\documentclass{article}

\usepackage{arxiv}

\usepackage[utf8]{inputenc} 
\usepackage[T1]{fontenc}    
\usepackage{hyperref}       
\usepackage{url}            
\usepackage{booktabs}       
\usepackage{amsfonts}       
\usepackage{nicefrac}       
\usepackage{microtype}      
\usepackage{lipsum}
\usepackage{graphicx}
\usepackage{amsmath}
\usepackage[square,numbers]{natbib}
\usepackage{siunitx}
\usepackage{caption}
\usepackage{subcaption}
\graphicspath{ {./images/} }

\title{A Bayesian Multisource Fusion Model for Spatiotemporal PM$_{2.5}$ in an Urban Setting}

\author{
 Abi I. Riley \\
  MRC Centre for Environment \& Health \\
  Department of Epidemiology and Biostatistics \\
  School of Public Health \\
  Faculty of Medicine \\
  Imperial College London\\ 
  United Kingdom \\
  W12 0BZ \\
  \texttt{a.riley21@imperial.ac.uk} \\
   \And
 Marta Blangiardo \\
  MRC Centre for Environment \& Health \\
  and UK Small Area Health Statistics Unit \\
  Department of Epidemiology and Biostatistics \\
  School of Public Health \\
  Faculty of Medicine \\
  Imperial College London\\ 
  United Kingdom \\
  W12 0BZ \\
  \And
 Frédéric B. Piel \\
  MRC Centre for Environment \& Health \\
  and UK Small Area Health Statistics Unit \\
  Department of Epidemiology and Biostatistics \\
  School of Public Health \\
  Faculty of Medicine \\
  Imperial College London\\ 
  United Kingdom \\
  W12 0BZ \\
  \And
  Andrew Beddows \\
  Environmental Research Group \\
  School of Public Health \\
  Faculty of Medicine \\
  Imperial College London \\
  United Kingdom \\
  W12 0BZ \\
  \And
  Sean Beevers \\
  Environmental Research Group \\
  School of Public Health \\
  Faculty of Medicine \\
  Imperial College London \\
  United Kingdom \\
  W12 0BZ \\
  \And
  Gary W. Fuller \\
  Environmental Research Group \\
  School of Public Health \\
  Faculty of Medicine \\
  Imperial College London \\
  United Kingdom \\
  W12 0BZ \\
  \And
  Paul Agnew \\
  Met Office \\
  Fitzroy Road \\
  Exeter\\
  United Kingdom\\
  EX1 3PB \\
  \And
  Monica Pirani \\
  MRC Centre for Environment \& Health \\
  Department of Epidemiology and Biostatistics \\
  School of Public Health \\
  Faculty of Medicine \\
  Imperial College London\\ 
  United Kingdom \\
  W12 0BZ \\
}

\begin{document}

\maketitle

\begin{abstract}
{Airborne particulate matter (PM$_{2.5}$) is a major public health concern in urban environments, where population density and emission sources exacerbate exposure risks. We present a novel Bayesian spatiotemporal fusion model to estimate monthly PM$_{2.5}$ concentrations over Greater London (2014-2019) at 1km resolution. The model integrates multiple PM$_{2.5}$ data sources, including outputs from two atmospheric air quality dispersion models and predictive variables, such as vegetation and satellite aerosol optical depth, while explicitly modelling a latent spatiotemporal field. Spatial misalignment of the data is addressed through an upscaling approach to predict across the entire area. 
Building on stochastic partial differential equations (SPDE) within the integrated nested Laplace approximations (INLA) framework, our method introduces spatially- and temporally-varying coefficients to flexibly calibrate datasets and capture fine-scale variability.
Model performance and complexity are balanced using predictive metrics such as the predictive model choice criterion and thorough cross-validation. The best performing model shows excellent fit and solid predictive performance, enabling reliable high-resolution spatiotemporal mapping of PM$_{2.5}$ concentrations with the associated uncertainty.
Furthermore, the model outputs, including full posterior predictive distributions, can be used to map exceedance probabilities of regulatory thresholds, supporting air quality management and targeted interventions in vulnerable urban areas, as well as providing refined exposure estimates of PM$_{2.5}$ for epidemiological applications.}
\end{abstract}

\keywords{Bayesian modelling, Air Pollution, Exposure Modelling, Spatio-temporal, Multisource}

\section{Introduction} \label{Section: Intro}
Decades of research in environmental sciences and public health have consistently identified pollution as one of the leading causes of excess morbidity and mortality and a contributor to global warming and environmental crises \cite{Alahmad2023ConnectionsHealth, Pozzer2023MortalityEstimates}. Air pollution alone was attributed to 8.1 million global deaths in 2021 by the Health Effects Institute \cite{HealthEffectsInstitute2024State2024}, with airborne particulate matter (PM) being identified as one of the top \emph{`major health-damaging air pollutants`} \cite{WorldHealthOrganization2022AmbientPollution}. 

PM with a diameter of 2.5 micrometres ($\mu$m) or less (PM$_{2.5}$) was the key focus of this study. This is based on mounting evidence of the adverse exposure effects of this pollutant \cite{WorldHealthOrganization2021WHOMonoxide}, including harmful and carcinogenic effects \cite{Turner2020OutdoorRecommendations}. PM$_{2.5}$ is a heterogeneous mixture of solids and aerosols, consisting of tiny liquid droplets, dry solid fragments, and solid cores coated with liquid layers, primarily emitted from domestic combustion, road transport, and industrial processes \cite{UKGovernment2019Clean2019}. This also includes sea salt, and desert dust especially from the Sahara desert over Europe, and secondary components are the results of oxidation reactions and other pollutants \cite{DepartmentforEnvironmentFoodRuralAffairs2023Air2022}. 

Target 11.6 from the 2030 United Nations Sustainable Development Goals (SDGs) aims to reduce the environmental impact of living in cities, by using measurements and reports of  \emph{`Annual mean levels of fine particulate matter (e.g. PM$_{2.5}$ and PM$_{10}$) in cities (population weighted)`} \cite{UnitedNationsTransformingDevelopment}. Where SDG Target 3.9 looks further to  \emph{`substantially reduce the number of deaths and illnesses from hazardous chemicals and air, water and soil pollution and contamination`} \cite{UnitedNationsTransformingDevelopment}. These key goals are measurable and long-term monitoring and assessment of air pollution levels and exposure, especially within urban environments \cite{Vilcassim2023GapsPollution}. 

The World Health Organization (WHO) and the UK have issued air quality targets to address these global aims \cite{WorldHealthOrganization2021WHOGuidelines,DepartmentforEnvironmentFoodRuralAffairs2010UKLimits}. Under the \textit{Environment Act 2021}, the UK aims to limit the annual mean concentration of PM$_{2.5}$ to \SI{10}{\micro\gram\per\cubic\meter} by 2040, while also reducing population exposure to PM$_{2.5}$ by 35\% compared to 2018 levels \cite{AirQualityExpertGroup2012FineKingdom}. Further, the WHO set the target of just  \SI{5}{\micro\gram\per\cubic\meter} as part of the 2021 air quality guidelines (AQGs) \cite{WorldHealthOrganization2021WHOGuidelines}. 

Accurate and comprehensive air pollution modelling can be used to monitor progress towards these aims and assess the environmental and medical effects of pollution concentrations. Relying solely on extrapolated data from sparse ground-based air pollution monitoring stations provides limited spatial coverage and accuracy and fails to capture fine spatial patterns, especially in urban areas \cite{DepartmentforEnvironmentFoodRuralAffairsMonitoringUK}. One effective approach is to use land-use regression (LUR) models, commonly incorporating spatially distributed geographical covariates into a linear regression model for the pollutant \cite{Hoek2008APollution}. Covariates are motivated by physical plausibility and previous work; PM$_{2.5}$ concentrations are intrinsically linked to land use, meteorological variables, population counts, and satellite-derived data \cite{deHoogh2018ModellingSwitzerland, Sahu2022BayesianR}. Developments of the LUR models explore using spatiotemporal data, proxy variables for air pollution measurements, and more complex model forms, as in \cite{Ma2024A2023}. However, \cite{Hoek2017MethodsPollutants} motivates the use of spatial and spatiotemporal modelling approaches with explicit underlying spatiotemporal model structures. These can also integrate different sources of air pollution data and predictive covariates and are often called data assimilation, data fusion, or data integration methods.

In this study, our goal was to develop a novel Bayesian spatiotemporal data fusion method to create a predictive map of PM$_{2.5}$ concentrations for Greater London (UK) from 2014 to 2019, providing monthly estimates at a high spatial resolution (1km x 1km grid). We used multiple sources of geospatial data and carefully considered the model structure to incorporate them, particularly calibrating numerical model outputs of air pollution estimates into our model. The Bayesian framework used takes into account prior knowledge of the model processes and variables, as well as the uncertainty in these priors and the model parameters and outputs \cite{vandeSchoot2021BayesianModelling}.

One of the main challenges was harmonising these large and diverse spatiotemporal datasets with different spatial and temporal scales and resolutions. This misalignment is a wider challenge in spatial statistics known in the literature as the change-of-support problem, and includes the related modiﬁable areal unit problem (MAUP) \cite{Gelfand2001OnData, Gelfand2012, Gryparis2009MeasurementEpidemiology}. Statistical data assimilation methods can be used to combine these different scale sources of air pollution data, model-derived data, and additional model predictors. \cite{Beroccal2019DataAssimilation} highlights two main statistical approaches: joint modelling and regression-based methods.

Joint modelling defines a shared underlying latent process between the observed values of the outcome and the alternative model variable. This can be powerful for understanding the 'true' spatiotemporal patterns of the pollutant and quantifying potential measurement error in the model. However, joint modelling methods are usually more computationally expensive, requiring simultaneous model running with a potentially complex joint process term \cite{Berrocal2011Space-TimeQuality}. In contrast, regression-based approaches more simply include the additional data source as a covariate in a regression model; however, this must be done carefully. Regression-based approaches are clearly less computationally intensive and can readily use power from alternative predictors of air pollution, like meteorology and environment types \cite{Sahu2022BayesianR}. 

Our Bayesian regression fusion model addressed the classic change-of-support problem by integrating point-referenced monitoring data with gridded satellite and model-derived predictors. The process involved upscaling individual station observations to a continuous spatial surface at a fixed 1 km × 1 km resolution, ensuring consistency across datasets. We additionally considered dynamic spatially- or temporally-coefficients for the satellite and model data as an additional calibration approach, as in \cite{Schliep2015AutoregressiveAOT}. The R package \texttt{INLA} provides a framework to approximate the full Bayesian inference of these models. \texttt{INLA} is named after the use of integrated nested Laplace approximations (INLA) \cite{Rue2009ApproximateApproximations}, and is a deterministic and fast alternative to the Markov chain Monte Carlo (MCMC) methods \cite{Hastings1970MonteApplications,Andrieu2003AnLearning}. The package is available from the R-INLA Project website (www.r-inla.org).

The proposed methodology relies on new processed data from a range of open-source products, including UK air pollution ground-monitoring data, global satellite-derived products, modelled gridded population and meteorological variables, and country-level land cover classification. The methodology consolidates and expands advanced modelling methods in R-INLA coupled with the stochastic partial differential equations (SPDE) approach \cite{Lindgren2011AnApproach}, which aims to capture physical diffusion mechanics.

We demonstrated the accuracy of our data fusion model through a rigorous cross-validation scheme. This finally widens the perspective of modelling air pollution in other urban areas with sparse ground monitoring data, by using open-access datasets \cite{Duncan2014SatelliteAvoid, Militino2018AnGeostatisticians}. 

The remainder of our paper is organised as follows. In Section \ref{Section: Data}, we outline the study domain and data sources. Section \ref{Section: Methods} covers the modelling methodology and parameters, followed by Section \ref{Section: Results} describing the results, model comparison, and cross-validation and finishing with the discussion and conclusions (Sections \ref{Section: Discussion} and \ref{Section: Conclusions}), where we critically examine the results and the significance of this work.

\section{Data}\label{Section: Data}
\subsection{Study Domain}
The spatial domain of this study was the highly populated and highly urbanised area of the Greater London Authority (GLA), as defined by the Office for National Statistics (ONS) \cite{GreaterLondonAuthority2019LondonDatastore}. This domain has a surface area of 1,569 km$^2$ with an estimated mid-2021 population of 8.797 million \cite{GreaterLondonAuthority2021LondonsPopulation}. The area is predominantly urban, lying between 51\textdegree N and 52\textdegree N, and is classified as a humid temperate oceanic climate by \cite{Beck2018PresentResolution}.

For this study, we performed predictions of fine air particles over a 1km x 1km spatial resolution grid encompassing the entire GLA area. 

This study's temporal domain covered January 2014 to December 2019, inclusive, with a resolution of calendar month.

\subsection{Monitoring site data}
In the GLA, the primary air pollution monitoring networks for PM$_{2.5}$ are the Automatic Urban and Rural Network (AURN) and the London Air Quality Network (LAQN) \cite{DepartmentforEnvironmentFoodRuralAffairsLocally-managedMonitoring}. The AURN is the leading UK automatic monitoring network \cite{DepartmentforEnvironmentFoodRuralAffairsUKResource} and complies with the EU Ambient Air Quality Directives \cite{EuropeanUnion2008DIRECTIVEEurope} for national reporting and monitoring purposes. The LAQN is an independent network provided by Imperial College London \cite{LondonAirLondonNetwork}, owned and funded by local authorities, and has traceability to national meteorological standards \cite{Mittal2023London2022}. 

Summary statistics for PM$_{2.5}$ measurements for London are freely available to download. The monthly mean concentrations were calculated for each monitoring site for each calendar month only if over 75\% of the daily measurements are available for that month. The data was downloaded from both sources (AURN: \url{uk-air.defra.gov.uk}, LAQN: \url{londonair.org.uk}). 

\begin{figure}
    \begin{subfigure}[b]{0.49\textwidth}
        \raggedleft
        \includegraphics[width=1\textwidth]{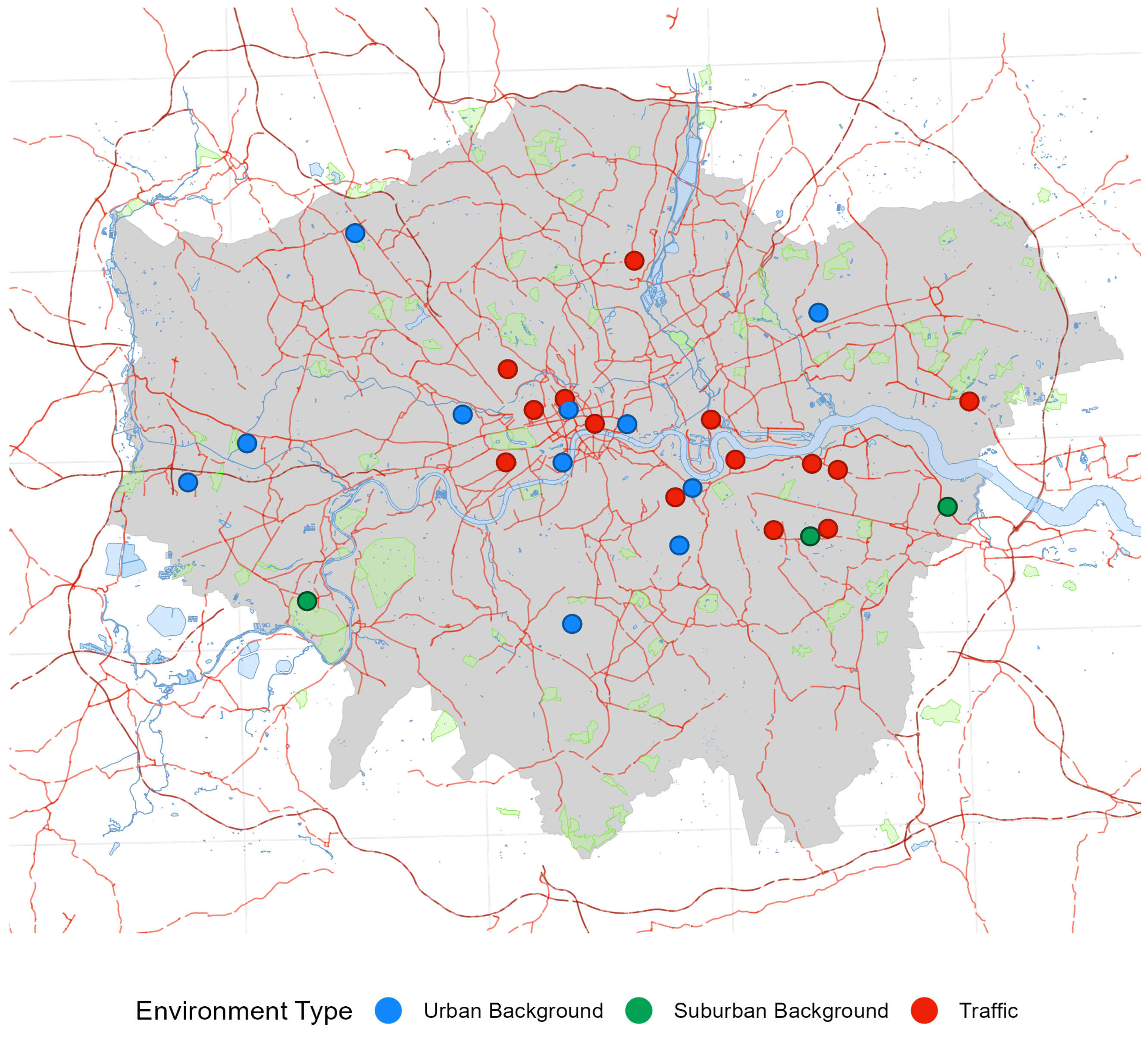}
         \caption{Map of the Greater London Authority Area (in gray) Automatic Urban and Rural Network (AURN) and the London Air Quality Network (LAQN) air pollution monitoring sites for PM$_{2.5}$, with major roads and landmarks from OpenStreetMap.\label{Fig: Monitoring Site Locations}}
    \end{subfigure}
    \begin{subfigure}[b]{0.49\textwidth}
        \raggedleft
        \includegraphics[width=1\textwidth]{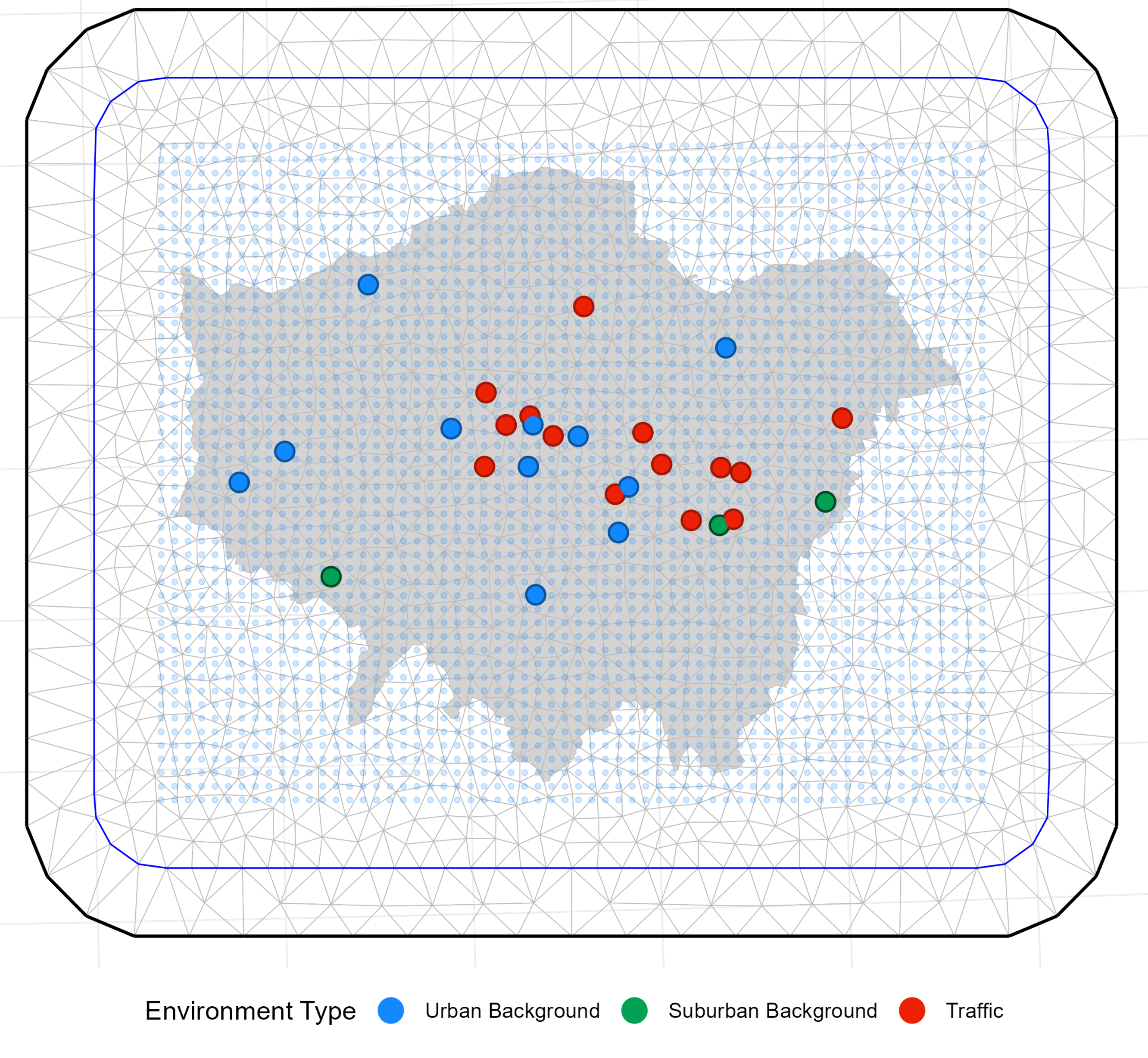}
        \caption{Map of INLA-SPDE mesh based on monitoring site locations and extent of the grid for prediction. Maximum edges set at 2.5km and 4km, for inner and outer mesh, respectively. \label{Fig: Mesh}}
    \end{subfigure}
    \caption{Mappings of monitoring site data and INLA mesh. \label{Fig: Map and Mesh}}
\end{figure}

PM$_{2.5}$ data was available for 44 monitoring sites (AURN — 11 sites, London Air — 33 sites) in the GLA Area. Each monitoring site was classified by its surrounding environment, and we consolidated these categories as urban background (17 sites), suburban background (2 sites), industrial (4 sites), and traffic (17 sites). Figure \ref{Fig: Monitoring Site Locations} shows the geographic location of the monitoring sites and the assigned environment type. Urban and suburban background areas are distinguished by the density of contiguous built-up area near the site; rural site types were not found within the study domain. Traffic was specified as either a site within either 5 or 10m of the kerbside for London Air and AURN sites, respectively. In line with similar UK-based works, we did not consider industrial sites in our models, because the model was primarily optimised for particle exposure estimation for epidemiological studies requiring knowledge of exposure at residential addresses.\cite{Pirani2014BayesianAreas,  Mukhopadhyay2018AWales, Wang2022PredictingBritain}. This exclusion also reflects the heterogeneity of industrial processes and emitted pollutants, as well as the limited spatial representativeness of point-level industrial monitoring within the 1 km$^2$ prediction grid \cite{Marshall2023OptionsReport}.

\subsection{Satellite-derived data}
Daily Aerosol Optical Depth (AOD) was obtained from NASA's multi-angle implementation of atmospheric correction (MAIAC) algorithm-based product (MCD19A2) \cite{Lyapustin2023MODIS/Terra+AquaSet}. This is a Level 2 reanalysis product derived from NASA's Moderate Resolution Imaging Spectroradiometer (MODIS) instruments aboard the Terra and Aqua satellites, and uses a time-series sliding window technique, utilising details about the cloud cover, land surface, and past observations to formulate and calibrate a global surface at a 1km x 1km spatial resolution \cite{NASA2022Multi-AngleMCD19}. 

The potential inclusion of this dataset was motivated by the correlation to ground-monitored PM$_{2.5}$ in similar international studies, e.g., \cite{He2021TheAOD, Pedde2022EstimatingData}.

Boundary Layer Height (BLH) is the altitude of the lowest layer of the atmosphere in metres. While not a direct predictor for PM$_{2.5}$, it was used as an adjustment of satellite-derived AOD, as seen in \cite{He2019AGuangzhou}. The monthly mean data is available from the European Centre for Medium-Range Weather Forecasts (ECMWF) Reanalysis v5 (ERA5), at a 0.25\textdegree x 0.25\textdegree \hspace{2mm} spatial resolution \cite{MunozSabater2019ERA5-LandPresent}.

\subsection{Pollution Climate Mapping (PCM) Model}
The Department for Environment, Food \& Rural Affairs (Defra)'s  Pollution Climate Mapping (PCM) model is a background concentration model for the UK, providing annual mean concentration surfaces on a 1km $\times$ 1km grid for PM$_{2.5}$ \cite{DepartmentforEnvironmentFoodRuralAffairsModelledData}. The PCM model for PM$_{2.5}$ is estimated through each constituent part, e.g. using measurements of sulphates, nitrates, and ammonium, employing atmospheric-dispersion models, and using estimated vehicle activity. Further details of the modelling methods used are available from \cite{RicardoEnergyEnvironment2021Technical2019}. The PCM was selected based on complementary work in the GLA, e.g., \cite{Forlani2020AR-INLA}. 

\subsection{UK Air Quality Reanalysis (AQR)}
UK Air Quality Reanalysis (AQR) dataset from the Met Office is a newly available output based on an underlying air quality model and post-processing for bias correction relative to observed ground monitoring observations \cite{MetOfficeUKReanalysis}. The underlying model, as also seen in \cite{Forlani2020AR-INLA} and other UK-based studies \cite{Mynard2023Long-termEvaluation}, is the Air Quality in the Unified Model (AQUM) \cite{UnifiedOffice, Savage2013AirEvaluation}. This model is based on several emission inventories and boundary conditions, using data from background labelled stations. Then, the post-processing is described in \cite{Neal2014ApplicationForecast}, which produces the final reanalysis data, available for the time hourly for 2003 to 2019, at a spatial resolution of 0.1\textdegree (roughly 10km). We used the readily available 0.1\textdegree resolution, monthly dataset from the Met Office Data Portal (\url{climatedataportal.metoffice.gov.uk}).

\subsection{Normalised Difference Vegetation Index (NDVI)}
Similarly to \citet{Hough2021GaussianFrance}, we included NDVI as a model predictor, as vegetation may affect dispersion, act as a natural filter and absorb some PM$_{2.5}$ from the air \cite{Ai2023TheChina}. NDVI quantifies the existence and density of vegetation and is calculated from the spectral reflectance at the red band ($ \textrm{Red} $) and the near-infrared bands ($ \textrm{NIR}$):

\begin{align} 
    \textrm{NDVI} &= \frac{\textrm{NIR} - \textrm{Red}}{\textrm{NIR} + \textrm{Red}}
\end{align}

The MODIS vegetation indices (MOD13A3) Version 6 provides monthly values of NDVI, with values from -1 to 1, for 1km x 1km grid cells; available from the NASA EarthData Store \cite{Didan2015MOD13A3Set}.

\subsection{Additional Predictive Variables}
Based on previous studies and physical processes, we considered a range of predictive variables to capture the dynamics of air particles. For example, \cite{Liu2009EstimatingInformation} shows the importance of seasonality and changing weather in predicting PM, while \cite{Huang2021ImpactAreas} discusses land use and cover in urban areas and the physical dynamics of air pollution. Therefore, we collected meteorological variables, including temperature, humidity, and precipitation, and physical geography-based variables, such as greenspace coverage, land cover, population density, and road density.

The meteorological variables were obtained from the HadUK-Grid from the Met Office \cite{Hollis2019HadUK-GridAObservations}, monthly at a spatial resolution of 1km x 1km. Greenspace can be described in two ways, NDVI, as above, or by the measured area of greenspace (including public parks, gardens, and playing fields) sourced from Ordnance Survey (OS) Data Hub \cite{OrdnanceSurvey2024OSGreenspace}. Land cover was obtained from the UK Environmental Information Data Centre (UK CEH), a 3-level land cover map \cite{Cole2021CorineGuernsey}, while population density was obtained from the WorldPop project \cite{Tatem2017WorldPopDemography}. Road density came from the OpenStreetMap road network, and was calculated as kilometres of road per grid cell.

Further details on these data sources are presented in the Supplementary Material and GitHub repository (\url{github.com/abiril/AirPollutionModel}).

\section{Methods}\label{Section: Methods}
\subsection{Modelling} \label{Subsection: Modelling}
The model for ambient PM$_{2.5}$ was defined within the computational framework provided by the INLA approach to latent Gaussian models (LGMs) and implemented through the R \texttt{INLA} package. The full methodology behind these models is described in \cite{Rue2009ApproximateApproximations}, and some of the more recent features and updates are described in \cite{Martins2013BayesianFeatures} and \cite{VanNiekerk2023AINLA}. 

In geostatistics, the INLA framework effectively utilises stochastic partial differential equations (SPDE; \citet{Lindgren2011AnApproach}),
approximating a Gaussian process model with a Mat\'{e}rn covariance function as a Gaussian Markov random field (GMRF; \citet{Rue2005GaussianApplications}), which is a multivariate normal distribution characterised by a sparse precision matrix. The approximation employs a linear combination of basis functions, defined in a triangulated domain called a mesh \cite{Lindgren2011AnApproach, Lindgren2022AFields}. Details are provided in the next section.

\subsubsection{Model Formulation}
First, we defined the general framework of the Bayesian geostatistical fusion model. 

Let $Y(\textbf{s}, t)$ denote the log-concentration of PM$_{2.5}$ at location $\textbf{s}$, for point-referenced coordinates $\textbf{s} \in \mathbb{R}^2$ and at time $t \in \mathcal{D}$, where $\mathcal{D}$ is the discrete temporal domain for months $t = 1,...,T$ (note that the logarithmic scale is used as pollution data are non-negative and skewed to the right). The model structure is:

\begin{align} 
    \label{Eq: Outcome} Y(\textbf{s},t) &= \eta(\textbf{s},t) + \varepsilon(\textbf{s},t) \\  
    \label{Eq: Linear Predictor} \eta(\textbf{s},t) &=  \mu(\textbf{s},t)  + \omega(\textbf{s},t)
\end{align}

Where $\eta(\textbf{s},t)$ is the underlying 'true' concentrations' value of the pollutants and  $\varepsilon(\textbf{s},t) \sim N(0, \sigma^2_\varepsilon)$ is a Gaussian white-noise measurement error term with nugget variance $\sigma^2_\varepsilon$. Then, we specified that $\eta(\textbf{s},t)$ is composed of a large-scale component, $\mu(\textbf{s},t)$, which captures the effect of the covariates, and a latent spatiotemporal process, $\omega(\textbf{s},t)$. 

We compared several different modelling formulations, making various modelling choices and assumptions for the forms of these modelling components. \\

\textbf{The baseline spatiotemporal model} \\ 
First, we considered a baseline stationary and isotropic spatiotemporal model using the INLA-SPDE approach as seen in several similar studies, including \cite{Cameletti2013Spatio-temporalApproach, Blangiardo2013SpatialR-INLA}. 

In particular, we assumed $\mu(\textbf{s},t)=\boldsymbol{\beta} \textbf{X}(\textbf{s},t)$, i.e., the large-scale component includes the linear effect of the covariates. We defined the spatiotemporal latent process term $\boldsymbol{\omega}$ as a temporally independent, mean-zero, stationary Gaussian spatiotemporal process with an underlying separable covariance form. The spatial process was defined as evolving in time as an autoregressive process with order 1 (AR1) and is specified as:
\begin{align} \label{Eq: omega}
    \omega(\textbf{s},t) = a \omega(\textbf{s}, t-1) + u(\textbf{s},t)
\end{align}
for $t = 1,...,T$, and coefficient $| a | < 1$, with $t = 1$ specified as:
\begin{align} \label{Eq: omega initial}
    \omega(\textbf{s}, 1) \sim N(0, \sigma^2_\omega/(1-a^2))
\end{align}

We then specified $u(\textbf{s},t)$ as a zero-mean Gaussian field, assumed to be temporally independent and characterised by the following spatiotemporal covariance function:

\begin{align} \label{Eq: Covariance}
    Cov(u(\textbf{s}_i, t_i), u(\textbf{s}_j, t_j)) =
    \begin{cases}
        Cov(u(\textbf{s}_i), u(\textbf{s}_j)), & \text{if} \hspace{1mm} t_i = t_j \\
        0, & \text{if} \hspace{1mm} t_i \neq t_j
    \end{cases} 
\end{align}
Here, the purely spatial covariance function is specified by a classic stationary  Mat\'{e}rn field:
\begin{align} \label{Eq: Matern}
    Cov(u(\textbf{s}_i), u(\textbf{s}_j)) &= \sigma^2_u \frac{2^{1-\nu}}{\Gamma(\nu)}\left( \kappa h \right) ^{\nu} K_{\nu}\left( \kappa h \right) \\
    \sigma_u^2 &= \frac{\Gamma(\nu)}{\Gamma(\alpha)4\pi\kappa^{2\nu} \tau^2}
\end{align}

where $h = ||\textbf{s}_i - \textbf{s}_j||$ is the absolute Euclidean distance between two spatial points $\textbf{s}_i, \textbf{s}_j \in \mathbb{R} ^2$, $\sigma_u^2$ is the marginal variance, $K_{\nu}$ is the 2$^{nh}$ order modified Bessel function with smoothing parameter, $\nu > 0$, and scaling parameter, $\kappa > 0$, which controls the spatial correlation range. The spatial range is calculated as $\rho = \frac{\sqrt{8\nu}}{\kappa}$ and is the approximate spatial distance where spatial correlation is 0.1.

To further detail the SPDE approach used to approximate this model, we first present the linear fractional stochastic partial differential equation (SPDE):

\begin{align}
    (\kappa^2 - \Delta)^{\alpha/2}(\tau \space u(\textbf{s}) ) = W(\textbf{s})
\end{align}

where $u(\textbf{s})$ is then a  Gaussian random field (GRF), $\kappa > 0$ is the scale parameter, $\alpha > d/2$ is the smoothness parameter for spatial dimension $d \in \mathbb{Z}$, $\Delta = \sum_{i=1}^d \frac{\partial^2}{\partial u^2_i}$ is the Laplacian operator, and $W(\textbf{s})$ is a Gaussian noise process. \cite{Lindgren2011AnApproach} shows that the exact solution to this SPDE is the stationary Gaussian field with the Mat\'{e}rn covariance, i.e., $u(\textbf{s})$, which links $\sigma_u^2$ and $\nu = \alpha - d/2$.  In detail, \cite{Lindgren2011AnApproach} in a purely spatial setting proposed a framework to represent spatial Mat\'{e}rn fields as GMRFs by discretising a diffusion-type SPDE (a stochastic generalisation of classical diffusion equations) using the Finite Element method on a triangulated mesh.

Then, there is a set of non-intersecting triangles with $G$ vertices, with corresponding deterministic basis functions, $\{\varphi_g\}$, with Gaussian mean-zero distributed weights, $\{\tilde{u}_g\}$.
\begin{align}
    u(\textbf{s}) = \sum^{G}_{g = 1} \varphi_g (\textbf{s}) \tilde{u}_g
\end{align}

Therefore, we write:
\begin{align} \label{Eq: GMRF}
    \tilde{u}(\textbf{s}, t) \sim N(0, \textbf{Q}_S^{-1})
\end{align}
where $\textbf{Q}_S$ is the precision matrix  for $\tilde{u}(\textbf{s},t)$ and linked to the  basis functions through:
\begin{align}
    \textbf{Q}_S = \tau( \kappa^2 \textbf{C} + 1\kappa^2\textbf{G} + \textbf{G}\textbf{C}^{-1}\textbf{G})
\end{align}
where $\textbf{C}$ is the diagonal matrix with $C_{ii} = \int \varphi_i(\textbf{s})d\textbf{s}$ and $\textbf{G}$ is the sparse matrix with $G_{ij} = \int \nabla\varphi_i(\textbf{s})d\textbf{s}\nabla \varphi_j(\textbf{s}) d\textbf{s}$.

For our application, we proposed the mesh in Figure \ref{Fig: Mesh}, with a fine maximum edge length of 2.5km and 4km, for the inner and outer meshes, respectively, and the boundaries determined by the full grid coordinates. 

The $G \times G$-dimensional precision matrix $\textbf{Q}_S$ is then a matrix of the correlations between the points along the vertices and is invariant in time. Using this, we redefine $\omega(\textbf{s},t)$ as:
\begin{align}
    \label{Eq: omega 2}\omega(\textbf{s},t) &= a \omega(\textbf{s}, t-1) + \tilde{u}(\textbf{s},t) \\
    \label{Eq: omega 2 initial}\omega(\textbf{s}, 1) &\sim N(0, \textbf{Q}^{-1})
\end{align}
with $\textbf{Q} = \textbf{Q}_T \otimes  \textbf{Q}_S$, where $\textbf{Q}_T$ is the precision matrix of a temporal AR1 process, as seen in \cite{Cameletti2013Spatio-temporalApproach}. This satisfies the separability assumption we made earlier for the spatiotemporal process, where the spatiotemporal covariance can be decomposed into separate spatial and temporal matrices.

\cite{Franco-Villoria2019AModels} discusses a general class of models called the \emph{`varying coefficient models`}, or VCMs, showing support for many applications. This work shows an increase in the model performance with no issues with overfitting, as expected with the increased flexibility of a model. However, it stresses the importance of priors, with penalised complexity (PC) priors being the suggested approach, and further discussion of priors is in \cite{Simpson2017PenalisingPriors}. We considered two main options: spatially-varying coefficients and temporally-varying coefficients.

\vspace{10pt}
\textbf{Spatially-varying coefficients} \\ 
To relax the spatial stationarity assumption above, we considered spatially-varying coefficients (SVC), allowing the marginal effects of the covariates to be spatially non-stationary \cite{Dambon2021MaximumPrediction, Ward2022IncorporatingModels}. In detail, we assumed that  $\mu(\textbf{s},t)=\boldsymbol{\beta}(\textbf{s}) \textbf{X}(\textbf{s},t)$. 

Based on the original idea in \cite{Gelfand2004NonstationaryCoregionalization}, we used the spatial SPDE model approach with a covariate-weighted projection matrix as in \cite[Chapter 8]{Krainski2018AdvancedINLA}. 

We specified the distribution of coefficient $\beta_k$ as a mean-zero, Gaussian distribution:
\begin{align} \label{Eq: SVC}
    \beta_k(\textbf{s}) \sim N(0, \sigma^2_{\beta} \Sigma)
\end{align}
with the spatial Mat\'{e}rn covariance function, $\Sigma$ as defined in Equation \ref{Eq: Matern}, weighted by coefficient $k$ and with spatial range $\rho_{\beta}$.

We evaluated the use of spatially-varying coefficients for the numerical model inputs, NDVI, and satellite-derived AOD, similarly to \cite{Franco-Villoria2019AModels}. This approach aimed to capture the finer climatic and atmospheric variations across London.

\vspace{10pt}
\textbf{Time-varying coefficients} \\
Time-varying coefficients (TVC) were considered in a similar way to the SVCs, but we specify a temporal model weighted by a covariate, which changes the assumption for the large-scale component, where  $\mu(\textbf{s},t)=\boldsymbol{\beta}(t) \textbf{X}(\textbf{s},t)$. We considered three simple specifications of a time model: identical and independent Normal (IID), AR1, and a random walk model of order 1 (RW1). 

We specify the distribution of the coefficient $\beta_k$ as the three different models:

\begin{align}\label{Eq:TVC}
    \text{TVC IID:} \hspace{10mm} &\beta_{k}(t) \sim N(0, \sigma_{\beta}^{2})  \\
    \text{TVC AR1:} \hspace{10mm} &\beta_{k}(1) \sim N(0,\sigma_{\beta}^{2}(1-\phi_{\beta}^2)^{-1}) \\
    & \beta_{k}(t) = \phi_\beta \beta_{k}(t-1) + \epsilon_t  \\
    & \epsilon_t \sim N(0, \sigma_{\beta}^{2}) \\
    \text{TVC RW1:} \hspace{10mm} &\beta_{k}(1) \sim N(0, \sigma_{\beta}^{2}) \\
    & \beta_{k}(t) - \beta_{k}(t-1) \sim N(0, \sigma_{\beta}^{2}) 
\end{align}

Where $t$ is the model time unit for each month ($t \in {1, ..., 72}$).

We evaluated TVC models for AQR output, NDVI and AOD, as these variables are highly seasonal and are known to chemically interact over time with components of PM$_{2.5}$. Similarly to \cite{Christopher2020GlobalRelationships}, we identified a complex relationship between PM$_{2.5}$ and variables such as AOD, with varying correlations over months, which a TVC aimed to capture.

\vspace{10pt}
\textbf{Summary of the Competitive Models} \\
Table~\ref{tab:All_Models} shows a summary of the considered competitive models. Three reference models are initially specified, namely a Bayesian hierarchical model that includes only additive, linear covariates and a model that includes solely the AR1 dynamic model for the latent process $\omega(\textbf{s}, t)$ with spatially correlated realisations $u(\textbf{s}, t)$. We then considered models of increasing complexity characterised by SVC and TVC terms.

\begin{table}[p]
    \centering
    \small\addtolength{\tabcolsep}{5pt}
    \begin{tabular}{l l}
        \hline
         Model & Model Terms\\
        \hline \hline
        Covariates Only &  $\eta(\textbf{s}, t) = \boldsymbol{\beta}\textbf{X}(\textbf{s}, t)$ \\
        \hline
        SPDE Only & $\eta(\textbf{s}, t) = \omega(\textbf{s}, t)$\\
        \hline \hline
        
        Baseline Model & $\eta(\textbf{s}, t) = \boldsymbol{\beta}\textbf{X}(\textbf{s}, t) + \omega(\textbf{s}, t)$  \\
        \hline 
        
        SVC Model for covariate $k$  &  $\eta(\textbf{s}, t) = \beta_{k}(\textbf{s})X_k(\textbf{s}, t) + \boldsymbol{\beta}\textbf{X}(\textbf{s}, t) + \omega(\textbf{s}, t)$  \\
        & $\beta_{k}(\textbf{s}) \sim N(0, \sigma^2_{\beta} \Sigma)$ \\
        
        \hline

        AR1 TVC Model for covariate $k$ & $\eta(\textbf{s}, t) = \beta_{k}(t)X_k(\textbf{s}, t) + \boldsymbol{\beta}\textbf{X}(\textbf{s}, t) + \omega(\textbf{s}, t)$ \\
        & $\beta_{k}(1) \sim N(0,\sigma_{\beta}^{2}(1-\phi_{\beta}^2)^{-1})$ \\
        & $\beta_{k}(t) = \phi_\beta \hspace{3pt} \beta_k(t-1) + \epsilon_t$ \\
        & $\epsilon_t \sim N(0, \sigma_{\beta}^{2})$ \\
        
        \hline
        
        RW1 TVC Model for covariate $k$ & $\eta(\textbf{s}, t) = \beta_{k}(t)X_k(\textbf{s}, t) + \boldsymbol{\beta}\textbf{X}_(\textbf{s}, t) + \omega(\textbf{s}, t)$ \\
        & $\beta_{k}(1) \sim N(0, \sigma_{\beta}^{2})$ \\
        & $\beta_{k}(t) - \beta_{k}(t-1) \sim N(0, \sigma_{\beta}^{2})$ \\
        \hline

        \hline \hline
        
    \end{tabular}  
    \caption{Summary of model term specifications. \label{tab:All_Models}}
\end{table}

\subsubsection{Model Prediction}
Given the observed values, $Y(\textbf{s}, t)$, for the set of monitoring locations $\textbf{s}$ and times $t = 1,...,T$, we rely on the posterior predictive distribution, similarly to \cite{Fasbender2009BayesianBelgium} and \cite{Cameletti2013Spatio-temporalApproach}, to obtain realisations $Y(\textbf{s}^*, t)$ for the new location of the grid $\textbf{s}^*$ at time $t$. Within the Bayesian framework, the posterior predictive distribution for new locations is obtained by integrating over the parameters with respect to the joint posterior distribution.

Therefore,  we evaluate the posterior predictive distribution for a set of new locations $\textbf{s}^*$ belonging to the regular 1km x 1km grid for each point in time $t = 1, ..., 72$ as:

\begin{align} \label{Eq: PPD}
    f(Y(\textbf{s}^*,t)|\textbf{Y}) = \int f(Y(\textbf{s}^*,t)|\boldsymbol{\theta}) f(\boldsymbol{\theta}|\textbf{Y}(\textbf{s},t))d\boldsymbol{\theta}
\end{align}

where $\textbf{Y}$ denotes all the observed data and $\boldsymbol{\theta}$ is the set of model parameters. Here, the product under the integral reduces to the joint posterior distribution  $f(Y(\textbf{s}^*,t)\boldsymbol{\theta}|\textbf{Y})$ and by marginalising out the parameters through the integral, we obtain the predictive distribution $f(Y(\textbf{s}^*,t)|\textbf{Y})$  of the concentrations of PM$_{2.5}$ at new locations at time $t$ given the measured data.

\subsubsection{Covariate Selection}
Covariate selection was performed prior to the addition of the spatiotemporal latent field, similarly to \cite{Maity2021BayesianSelection}. We evaluated the inclusion or exclusion of each covariate as a fixed linear effect in the simple Bayesian regression model: $Y = \boldsymbol{\beta} \textbf{X} + \varepsilon$.

In particular, variable selection was then performed through a stepwise selection method, as seen in \cite{Lu2022BayesianPerspectives}. We considered a wide range of covariates in the initial analysis, including mean temperature, humidity, precipitation, pressure, evapotranspiration, wind direction and speed, AOD, PCM, AQR, population count, land cover, NDVI, distance to roads and road density, and a site type indicator. All variables were standardised to have a mean of zero and a variance of one.

This process and the results are shown in the Supplementary Material. Final model variable selection was based on this and comparison of full models using the predictive model choice criterion (PMCC), as seen in \cite{Sahu2012HierarchicalModelling}. 

\subsubsection{Model Implementation}
All models were implemented using the PM$_{2.5}$ data presented in Figure \ref{fig: Monitoring Sites} and is all available from the GitHub repository (\url{github.com/abiril/AirPollutionModel}). 

In the construction of the mesh, we set the maximum inner edge at $2.5km$, which is less than $1/10$th of the variogram range, $50 km$, as recommended in the tutorial \cite{Belmont2022BuildingINLA}. Further, as the inner domain encapsulates all of our spatial area, we chose the outer extension with care to reducing boundary effects. In detail, we set the larger length of the edge to $4km$, and with an offset $5km$, this simple outer extension improved computation time and reduces complexity. Alternately to the default setting, we set the SPDE smoothness parameter as $\alpha = 3/2$, which simplifies the Mat\'{e}rn covariance function to the exponential form, as discussed in \cite{Lindgren2011AnApproach}.

In setting up the SPDE, we specified penalised complexity priors (PC-prior) \cite{Fuglstad2015ConstructingFields} for the range $\rho$ and marginal standard deviation $\sigma$ \cite{Lindgren2011AnApproach}: $Prob(\rho > \rho_0) = p $ where $\rho_0 = 20$, $p = 0.01$, and $Prob(\sigma > \sigma_0) = p $ where $ \sigma_0 = 0.1$, $p = 0.1$. Additionally, specifying a PC-prior for the AR1 temporal structure on the spatiotemporal term: $Prob(a > a_0) = \alpha$ where $a_0 = 0.95$ and $alpha = 0.5$. 

Based on preliminary variable selection, as described in the supplementary material, we primarily considered the set of additive linear predictors: the PCM and AQR models, the background indicator, and the surrounding environment terms for NDVI and road density. We used the default priors for the fixed effects, i.e., $\beta_i \sim N(0,1000)$.

As in Equation \ref{Eq: SVC}, we considered the use of the spatial SPDE approach with a covariate-weighted correlation matrix for the covariates: AOD, AQR and NDVI. On the spatial SPDE terms, we set the priors: $Prob(\rho_{\beta} > \rho_0) = p $ where $\rho_0 = 20$, $p = 0.01$, and $Prob(\sigma_{\beta} > \sigma_0) = p $ where $ \sigma_0 = 0.1$, $p = 0.1$. 

Next, we considered the additional time-varying coefficient (TVC) term for covariate $k$,\{AOD, AQR, NDVI, PCM\}, with either the unstructured IID model or one of AR1 or RW1 models. Priors and initial values can be set too for the precision of the TVC model ($\tau_{\beta} = 1 \slash\sigma^2_{\beta}$) and for AR1, the spatial range ($\rho_{\beta}$).

\subsubsection{Model Comparison} \label{Section: Model Comparison}
\paragraph*{Model Fit} \hfill \break
To assess and compare the fit of the proposed models, we used two statistics: the squared correlation coefficient, $R^2$, and the predictive model choice criterion (PMCC; \citet{Sahu2012HierarchicalModelling}). The correlation term is easily interpretable and is a common goodness-of-fit statistic used in many settings, including Bayesian spatiotemporal modelling \cite{Gelman2014UnderstandingModels}. In detail, letting the observed value, $Y(\textbf{s},t)$, be denoted by $y_{it}$ for monitoring site $\textbf{s}_i$, for $i = 1, \dots, n$ at time $t$, for $t = 1, \dots, T$, then the mean observed value is denoted by $\bar{y}$. The corresponding posterior fitted value is denoted by $y'_{it}$ with mean $\bar{y}'$.
\begin{align}
    R^2 = \frac{(\sum^{n}_{i = 1}\sum^{T}_{t = 1}{(y_{it} - \bar{y})(y'_{it} - \bar{y}'))^2}}{\sum^{n}_{i = 1}\sum^{T}_{t = 1}{(y_{it} - \bar{y}) \sum^{n}_{i = 1}\sum^{T}_{t = 1}}{(y'_{it} - \bar{y}')}}
\end{align}

We also considered the PMCC, as in \cite{Sahu2012HierarchicalModelling}, where, with the above notation, it is calculated by:

\begin{align}
    PMCC &= \sum^{n}_{i = 1}\sum^{T}_{t = 1}{(y_{it} - y'_{it})^2} + \sum^{n}_{i = 1}\sum^{T}_{t = 1}{Var(y'_{it})}
\end{align}

The first term in the above equation is a goodness-of-fit term, while the second is a predictive variance penalty term for model complexity. This statistic is first presented in \cite{Gelfand1998ModelApproach} and \cite{Ibrahim2001Criterion-BasedAssessment} and is derived from the posterior predictive distribution. It aims to assess the variability and precision of the posterior predictions at the monitoring sites, towards our modelling aim. The model with the smallest PMCC value is selected among the competing models.

\paragraph*{Predictive Performance} \label{Sec: Pred Performance} \hfill \break
To assess the predictive performance of the models, we performed spatial and temporal cross-validation studies. Cross-validation methods partition the observed dataset into training and testing sets, fitting the model to the training set and predicting it on the test set. Finally, computing statistics based on the differences between the observed testing dataset and predicted values at these specific points and times.

We considered both a spatial and a temporal cross-validation method, similar to \cite{Roberts2017Cross-validationStructure} and \cite{VandenBossche2020ACarbon}: 

\begin{itemize}
    \item \textbf{k-fold Temporal Cross-Validation}: For integer $k$, we divide the time period into $k$ equal periods and perform prediction for each time period using the other time periods as training set.
\end{itemize}
\hspace{8mm} We use $k = 6$ and partition by calendar year from 2014 to 2019.
    
\begin{itemize}
    \item \textbf{k-block Spatial Cross-Validation}: For integer $k$, we partition our study spatial domain into $k$ spatial blocks with an equal number of sites, then perform and assess prediction on each block in turn. 
\end{itemize}
\hspace{8mm} We use $k = 6$ and partition the London domain into 2 x 3 blocks -  \\
\hspace{10mm} North East, North, North West, South West, South, South East.

Denoting the predicted posterior value as $y^*_{it}$ for unobserved sites $i = 1,\dots,n$ and times $t = 1,\dots,T$, we calculated: the predicted R$^2$, the Root Mean Squared Error (RMSE), the Bias; and the the empirical coverage of 95\% predictive credible intervals (Cov). The R$^2$ is as above, while the RMSE and Bias are:

\begin{align}
    RMSE &= \sqrt{\frac{1}{nT} \sum^{n}_{i = 1}\sum^{T}_{t = 1}{(y_{it} - y^*_{it})^2}} \\
    Bias &= \frac{{\sum^{n}_{i = 1}\sum^{T}_{t = 1}{(y_{it} - y^*_{it})}}}{nT}\\
\end{align}

The empirical coverage of 95\% predictive credible intervals  \cite{McCandless2007BayesianStudies} is used to evaluate whether the predictive uncertainty is properly quantified. If the empirical coverage falls below the nominal level, assuming unbiased predictions, the model underestimates the variability in the predictions. Conversely, if the empirical coverage exceeds the nominal level, the model overestimates the predictive uncertainty.

\section{Results}\label{Section: Results}
\begin{figure}
    \includegraphics[width=1\textwidth]{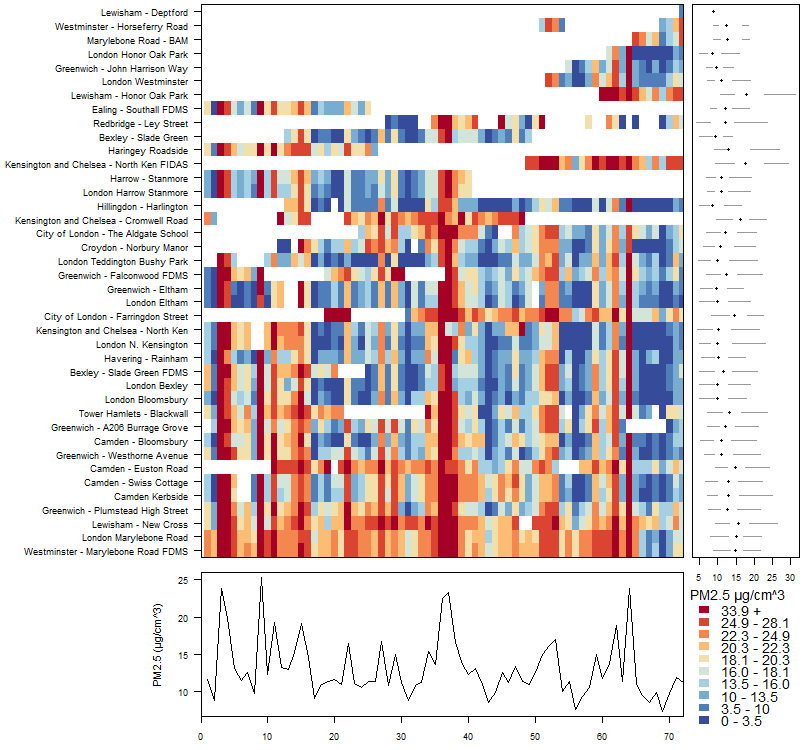}
    \caption{\textbf{Main}: Mean PM$_{2.5}$ concentrations in $\mu g/m^3$ by monitoring site (y-axis) across the 72 months (x-axis). \textbf{Right}: Box plot of mean PM$_{2.5}$ concentration in  $\mu g/m^3$ by monitoring site (y-axis). \textbf{Bottom}: Averaged time series of mean PM$_{2.5}$ concentrations in $\mu g/m^3$ across all monitoring sites across the 72 months (x-axis). Multivariate time series plot from R package \texttt{mvtsplot}. \label{fig: Monitoring Sites}}
\end{figure}

A summary of monthly monitoring site data is given in Figure \ref{fig: Monitoring Sites}. Through the study period, we observed an overall decrease in PM$_{2.5}$ concentrations and noticed a strong seasonality trend. This is in line with the data presented in the 'Air Pollution in the UK 2022' report \cite{DepartmentforEnvironmentFoodRuralAffairs2023Air2022}. After careful data selection and pre-processing, 71\% of the total monthly values were available and used in the model.

Table \ref{tab:Model Stats} shows the model fit statistics for the top ten competitive models. It includes the squared correlation coefficient, R$^2$, and the PMCC, calculated by comparing observed and posterior fitted values, in addition to the length of the computation time given in hours. 

\begin{table}[p]
    \centering
    \small\addtolength{\tabcolsep}{5pt}
    \begin{tabular}{ l | l | c c c  }
        \hline
        & Model Formula & \multicolumn{3}{c}{Model Fit} \\
        Model &  Formula for $\eta(\textbf{s},t)$ & R$^2$ & PMCC & Time \\ 
        
        \hline \hline
        Model 0a &  $ \beta_0 + \beta_1 BG(\textbf{s})  +  \beta_2 PCM(\textbf{s},t) + \beta_3 AQR(\textbf{s},t)$  & 0.82 & 403 & 0.03 \\
        
        Model 0b &  $ \beta_0 + \beta_1 BG(\textbf{s})  + \beta_2 PCM(\textbf{s},t) + \beta_3 AQR(\textbf{s},t)  +\beta_4 NDVI(\textbf{s},t)$ & 0.82 & 402 & 0.03 \\
        
        Model 0c & $ \beta_0 + \omega(\textbf{s},t)  $ & 0.94 & 203 & 2.69 \\

        \hline \hline

        Model 1a & $ \beta_0 + \beta_1 BG(\textbf{s}) + \beta_2 PCM(\textbf{s},t)   + \beta_3 AQR(\textbf{s},t) $ &  0.94 & 194 & 1.7 \\

        Model 1b & $ \beta_0 + \beta_1 BG(\textbf{s})  + \beta_2 PCM(\textbf{s},t) + \beta_3 AQR(\textbf{s},t)  +  \beta_4 NDVI(\textbf{s},t)$ &  0.94 & 193 & 1.55 \\

        Model 2a & $\beta_0 + \beta_1 BG(\textbf{s}) + \beta_2 PCM(\textbf{s},t) + \beta_3(\textbf{s}) AQR(\textbf{s}, t) +   \omega(\textbf{s},t) $ & 0.94 & 191 & 3.8 \\

        Model 2b & $ \beta_0 + \beta_1 BG(\textbf{s}) +  \beta_2 PCM(\textbf{s},t) + \beta_3 AQR(\textbf{s},t)  + \beta_4(\textbf{s}) NDVI(\textbf{s}, t) + \omega(\textbf{s},t) $ & 0.94 & 194 & 5.0 \\

        Model 3a & $\beta_0 + \beta_1 BG(\textbf{s}) + \beta_2 PCM(\textbf{s},t) + \beta_3(t) AQR(\textbf{s},t) + \omega(\textbf{s},t) $ & 0.94 & 189 & 3.4 \\
        & \hspace{5mm} $\beta_3(t) \sim AR(1)$ for time $t$ & & & \\

        Model 3b & $ \beta_0 + \beta_1 BG(\textbf{s}) + \beta_2 PCM(\textbf{s},t) + \beta_3 AQR(\textbf{s},t)  +  \beta_4(t) NDVI(\textbf{s},t) + \omega(\textbf{s},t) $ & 0.94 & 188 & 4.3 \\
        & \hspace{5mm} $\beta_4(t) \sim AR(1)$ for time $t$ & & & \\

        Model 4a &  $\beta_0 + \beta_1 BG(\textbf{s}) + \beta_2(\textbf{s}) PCM(\textbf{s},t) +  \beta_3(t)AQR(\textbf{s},t) +  \omega(\textbf{s},t)$ & 0.94 & 189 & 10.4 \\
        & \hspace{5mm} $\beta_3(t) \sim AR(1)$ for time $t$ & & & \\
        
        Model 4b & $\beta_0 + \beta_1 BG(\textbf{s})  + \beta_2 PCM(\textbf{s},t) + \beta_3(\textbf{s}) AQR(\textbf{s},t) + \beta_4(t) NDVI(\textbf{s},t) + \omega(\textbf{s},t)$ & \textbf{0.94} & \textbf{185} & 3.6 \\
        & \hspace{5mm} $\beta_4(t) \sim AR(1)$ for time $t$ & & & \\

        Model 4c & $\beta_0 + \beta_1 BG(\textbf{s})  + \beta_2(\textbf{s}) PCM(\textbf{s},t) + \beta_3(t)AQR(\textbf{s},t) + \beta_4(\textbf{s})NDVI(\textbf{s},t) +  \omega(\textbf{s},t)$ & 0.94 & 189 & 7.7 \\
        & \hspace{5mm} $\beta_3(t) \sim AR(1)$ for time $t$ & & & \\

        Model 4d & $\beta_0 + \beta_1 BG(\textbf{s})  + \beta_2(\textbf{s}) PCM(\textbf{s},t) + \beta_3(\textbf{s})AQR(\textbf{s},t) + \beta_4(t)NDVI(\textbf{s},t) +  \omega(\textbf{s},t)$ & 0.94 & 187 & 8.9\\
        & \hspace{5mm} $\beta_4(t) \sim AR(1)$ for time $t$ & & & \\

        \hline \hline
    \end{tabular}
    \caption{Model formulae and model fit statistics for the 3 baseline and 10 competitive models. R$^2$: squared correlation coefficient; PMCC: Predictive Model Choice Criterion; and Time: INLA computation time in hours.}
    \label{tab:Model Stats}
\end{table}

All models demonstrated a high and consistent goodness-of-fit, with R$^2$ values ranging from 0.941 to 0.943 and the PMCC between 185.31 and 194.87. 

The more complex models, characterised by varying coefficients (M2 – M4), outperformed the baseline model (M1), although this came at the cost of increased computational time, from approximately 1.74 hours for M1 to between 3.85 and 10.36 hours for M2 – M4.

\begin{table}[p]
    \centering
    \small\addtolength{\tabcolsep}{5pt}
    \begin{tabular}{ l | c c c c  | c c c c  }
        \hline
        & \multicolumn{4}{c|}{Temporal CV} & \multicolumn{4}{c}{Spatial CV} \\
        Model & R$^2$ & RMSE & Bias & Cov & R$^2$ & RMSE & Bias & Cov  \\ 
        \hline \hline
        Model 0a & 0.67 & 0.21 & 0.004 & 0.09 & 0.71 & 0.20 & 0.004 & 0.11 \\
        Model 0b & 0.67 & 0.21 & 0.001 & 0.10 & 0.71 & 0.20 & 0.003 & 0.11 \\
        Model 0c & 0.19 & 0.35 & 0.062  & 0.99 & 0.59 & 0.27 & 0.014  & 0.98 \\
        \hline \hline
        Model 1a & 0.71 & 0.19 & 0.003 & 0.89 & 0.69 & 0.22 & -0.024 & 0.91 \\
        Model 1b & 0.71 & 0.19 & 0.001 & 0.89 & 0.69 & 0.22 & -0.027 & 0.91 \\
        Model 2a & 0.72 & 0.19 & 0.004  & 0.89 & 0.69 & 0.22 & -0.023 & 0.91 \\
        Model 2b & 0.71 & 0.19 & 0.002  & 0.89 & 0.69 & 0.22 & -0.025 & 0.91 \\
        Model 3a & 0.72 & 0.19 & 0.012  & 0.90 & 0.70 & 0.21 & -0.027 & 0.91 \\
        Model 3b & 0.71 & 0.19 & -0.007 & 0.90 & 0.69 & 0.22 & -0.025 & 0.91 \\
        Model 4a & 0.71 & 0.20 & 0.015  & 0.90 & 0.70 & 0.21 & -0.029 & 0.91 \\
        Model 4b & 0.72 & 0.19 & -0.007 & 0.90 & 0.68 & 0.22 & -0.024 & 0.91 \\
        Model 4c & 0.71 & 0.20 & 0.013  & 0.90 & 0.70 & 0.21 & -0.031 & 0.91 \\
        Model 4d & 0.71 & 0.20 & -0.006 & 0.89 & 0.68 & 0.22 & -0.027 & 0.92 \\
        \hline \hline
    \end{tabular}
    \caption{Model predictive performance statistics for the three reference and ten comparative models. For temporal cross-validation and spatial block cross-validation: (1) \textbf{R$^2$} the squared correlation coefficient between observed and predicted, (2) the \textbf{RMSE}, (3) the \textbf{Bias}, and (4) the empirical coverage of 95\% CI, \textbf{Cov}. All values shown on the log-scale. \label{tab:Model Pred}}
\end{table}

Table \ref{tab:Model Pred} presents the performance metrics for the evaluation of the predictive capability of the baseline and competitive models: (1) the squared correlation coefficient between observed and predicted, $R^2$, (2) the RMSE, (3) the bias, and (4) the empirical coverage of 95\% CI, Cov, on the log-scale, for both the temporal cross-validation and spatial block cross-validation.

First, the covariate-only models (Model 0a and Model 0b) and the SPDE-only model (Model 0c) both showed a moderate ability to fit to the data and capture the observed spatiotemporal trends. Using the carefully chosen covariates, Model 0a and Model 0b ran quickly and achieved good predictive squared correlation coefficients in cross-validation (Temporal CV R$^2$: 0.667 and 0.666; Spatial CV R$^2$: 0.708 and 0.710) but have a lower coverage level. In contrast, Model 0c has a very good model fit but exhibits a larger coverage rate than the nominal value, 0.95 (Temporal CV Cov: 0.989; Spatial CV Cov: 0.977). 

The baseline models (Model 1a and Model 1b) were built on the reference models, using added model covariates to capture most of the spatiotemporal variation in PM$_{2.5}$ and the SPDE term to capture the residual underlying process, as conceptualised in \cite{He2024Spatio-temporalApproach}. This model demonstrated significant improvements over the reference models, including enhanced model fit (R$^2$), improved overall temporal cross-validation metrics, and better spatial cross-validation coverage. Model 1b included an intercept and four model covariates, with posterior estimated mean (95\% credible interval in brackets) of the covariate coefficients: Intercept: \textbf{2.60 (2.50; 2.60)}; PCM: 0.00 (-0.03; 0.03); AQR: \textbf{0.28 (0.27; 0.29)}; BG: \textbf{-0.27 (-0.32;-0.27)}; and, NDVI: -0.04 (-0.14; 0.07). 

From the proposed options for SVCs, both Model 2a and 2b had good R$^2$ fit values but relatively high PMCC (Model 2a: 191; Model 2b: 195). Model 2a performed especially well in temporal CV metrics R$^2$ and RMSE, and spatial CV bias; which is balanced with low temporal CV bias and coverage. Similarly, Models Model 3a and Model 3b use AQR and NDVI, respectively, as variables with TVCs. The prevailing model choice here used time in months ($t = 1,...,72$) and the AR1 model term; IID models had low model fit and RW(1) models increased computational time. However, these AR1-based models took 3.4 and 4.2 hours, and the PMCCs were lower than Models 1 and 2.

Based on the results on Models 2 and 3, we considered 4 combinations of model terms. Model 4a: SVC PCM and TVC AQR; Model 4b: SVC AQR and TVC NDVI; Model 4c: SVC PCM \& NDVI and TVC AQR; and, Model 4d: SVC PCM \& AQR and TVC NDVI. Models 4a, 4c and 4d have a marked increase in computational time (Model 4a: 10.4 hours; Model 4c: 7.7 hours; Model 4d: 8.9 hours). However, Model 4b had the overall best model fit with R$^2$ = 0.94 and PMCC = 185, and a computation time of just 3.6 hours. 

\begin{figure}
    \includegraphics[width=0.8\textwidth]{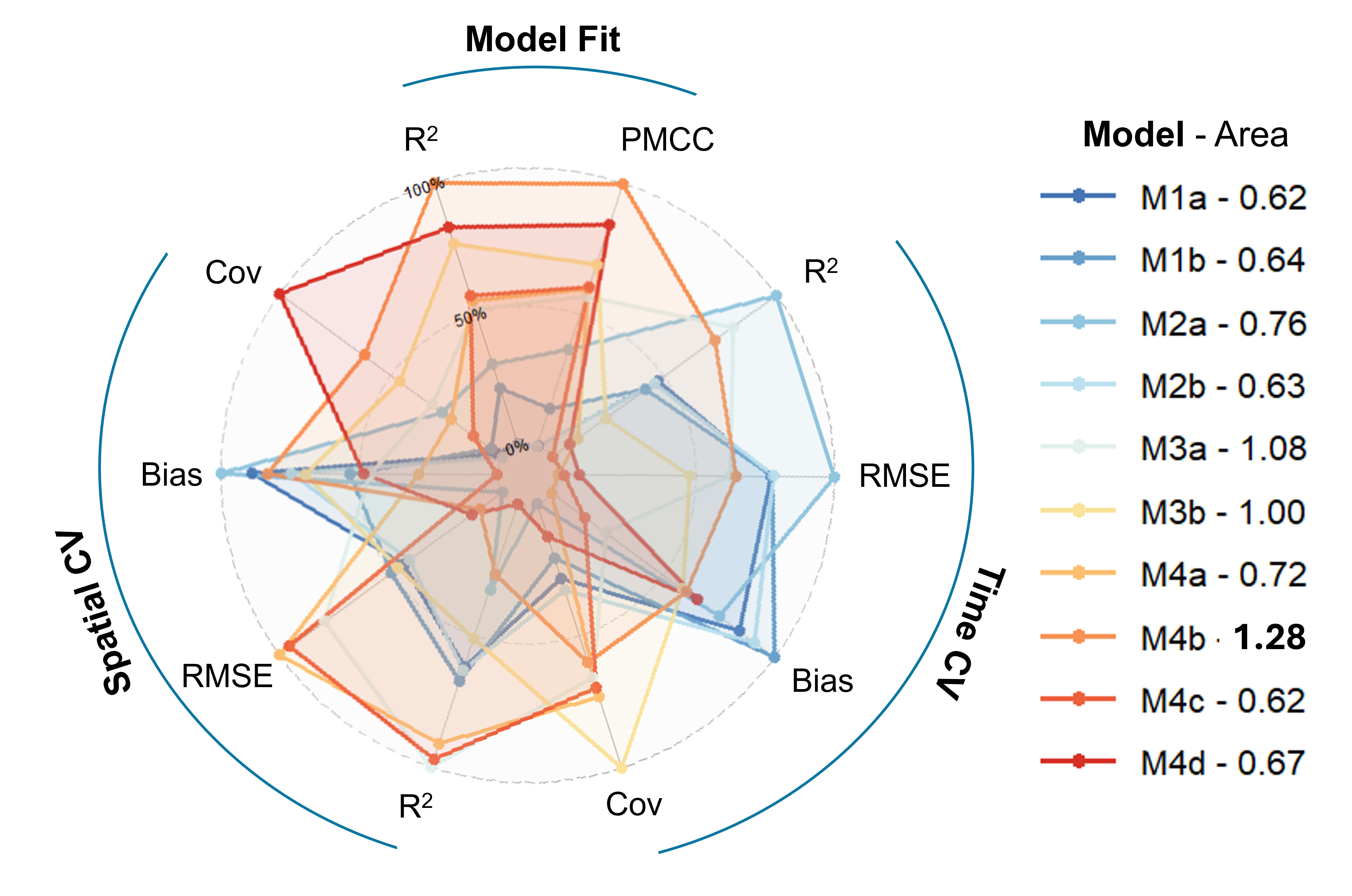}
    \caption{Radar plot of model fit and predictive performance metrics for the competing models. Metrics are rescaled to a 0–100\% scale, where 100\% corresponds to the optimal value for each statistic. The area of each radar shape summarises the overall performance of the model. \label{Fig: Radar}}
\end{figure}

We visualised and compared the model performance using a radar plot in Figure \ref{Fig: Radar}. The model fit and CV statistics were scaled from 0 to 100\%, where 100\% represents the optimal value for each metric (e.g., highest R$^2$, or coverage closest to 95\%). To support the comparison, we also calculated the area of each radar shape, with larger areas indicating better overall performance.

Based on this model selection strategy, we selected Model 4b, characterised by SVC for the AQR variable, TVC for NDVI modelled using an AR1 specification, linear effect for PCM's output, the background indicator for site type, and the latent spatiotemporal process. This model represents an overall compromise of the model fit and CV metrics.

This chosen 'best' model was then re-run with settings to produce more accurate approximations and improve performance. In line with \cite{Martino2019IntegratedINLA} and \cite{Rue2017BayesianReview}, we selected the full 'Laplace' approximation method. This increased computational time significantly from 3.6 hours to 5.0 hours. This final model was run on a high-performance computing cluster using 96GB of memory across 10 cores.

Examining the cross-validation results, we can report the individual performance measures for the study years and spatial block areas (Table \ref{tab:Cross-Val Details}). Across the model years, there is a slight pattern for model performance, where 2014 has particularly good performance (highest $R^2$ = 0.84, smallest RMSE =  0.16) and it decreases over time, with 2019: R$^2$ = 0.70, RMSE = 0.22, and Bias = 0.09. Whereas 2019 has a high RMSE and bias, but the most optimal value for coverage. 

For the spatial block cross-validation, we see the best metrics for North and South West (highest R$^2$ = 0.91, 0.77; lowest RMSE = 0.18, 0.14). However, these areas have near-perfect coverage, well beyond the nominal value. South generally performs worst, with the lowest R$^2$, highest RMSE, and lowest coverage.

\begin{table}[p]
    \centering
    \small\addtolength{\tabcolsep}{5pt}

    \begin{tabular}{lllll}
        \hline
         Model &R$^2$ & RMSE & Bias & Cov \\
         \hline \hline         
         
         \multicolumn{5}{l}{\textbf{Years}} \\
         \hline
         2014 & 0.85 & 0.16 & -0.05 & 0.98 \\
         2015 & 0.65 & 0.20 & -0.03 & 0.89 \\
         2016 & 0.76 & 0.18 & -0.03 & 0.87 \\
         2017 & 0.69 & 0.19 & -0.01 & 0.84 \\
         2018 & 0.66 & 0.19 & -0.005 & 0.89 \\
         2019 & 0.70 & 0.23 & 0.09 & 0.91 \\
         
         \hline \hline
         \multicolumn{5}{l}{\textbf{Spatial Blocks}} \\
         \hline
         South West & 0.77 & 0.18 & 0.05 & 1.00 \\ 
         North West & 0.91 & 0.14 & -0.09 & 1.00 \\ 
         South & 0.52 & 0.27 & -0.02 & 0.82 \\
         North & 0.63 & 0.22 & -0.05 & 0.94 \\
         South East & 0.71 & 0.25 & -0.17 & 0.79 \\
         North East & 0.57 & 0.26 & 0.13 & 0.95 \\
         \hline 

    \end{tabular}
    \caption{Cross-validation prediction performance metrics for individual spatial and temporal folds. \label{tab:Cross-Val Details}}
   
\end{table}

\begin{figure}
    \includegraphics[width=0.9\textwidth]{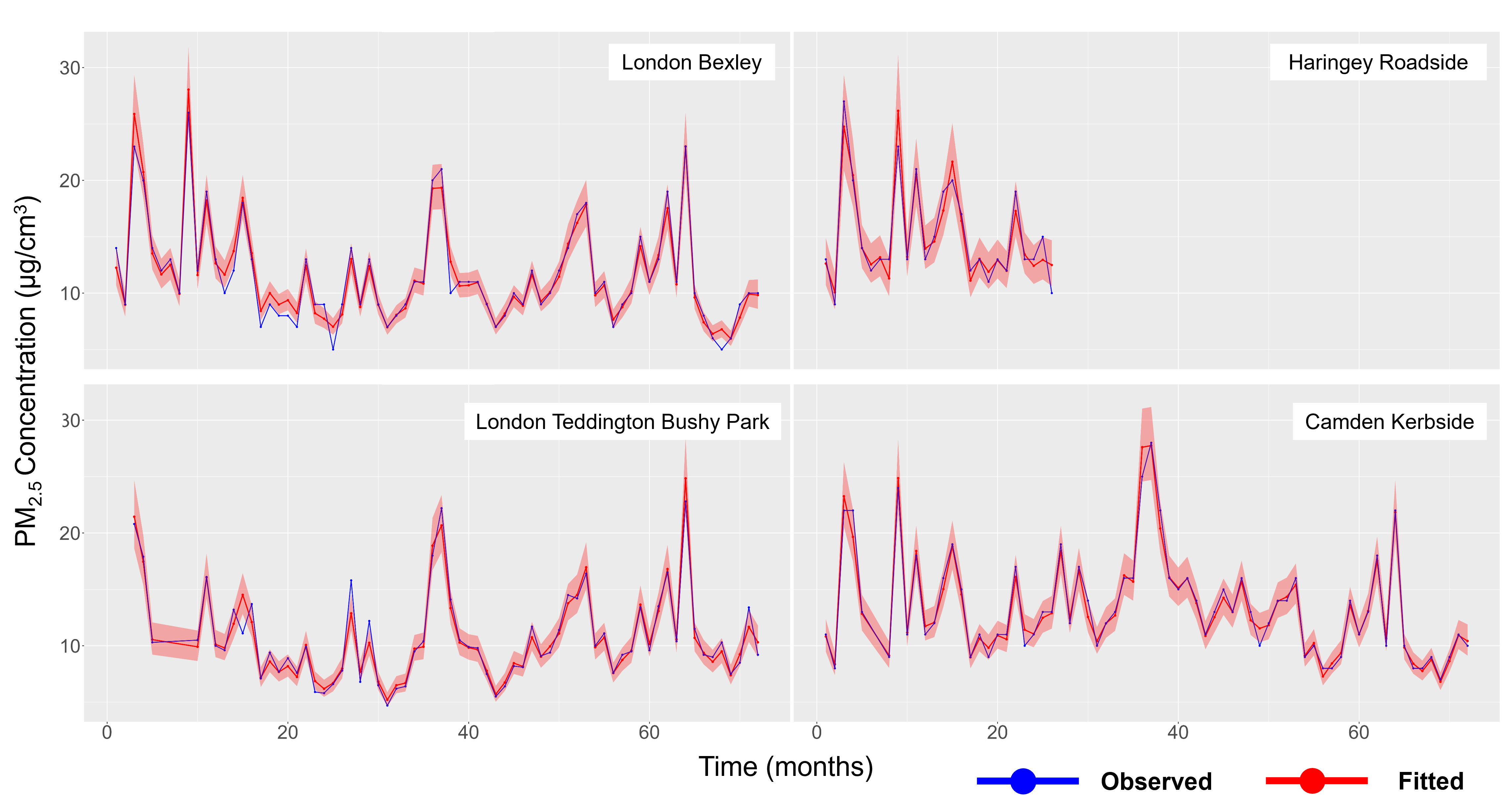}
    \caption{Observed monthly mean PM$_{2.5}$ concentrations versus posterior fitted PM$_{2.5}$ values with 2.5\% and 97.5\% credible intervals at four example monitoring sites in Greater London. London Bexley (AURN; background site), Haringey Roadside (AURN; traffic site), London Teddington Bushy Park (London Air; background site), Camden Kerbside (London Air; traffic site).  \label{Fig: Obvs vs Fit}}
\end{figure}

Figure \ref{Fig: Obvs vs Fit} shows a sample of four London monitoring stations, comparing observed PM$_{2.5}$ concentrations with the fitted values from Model 4b and their associated uncertainty. 

\begin{figure}
    \includegraphics[width=1\textwidth]{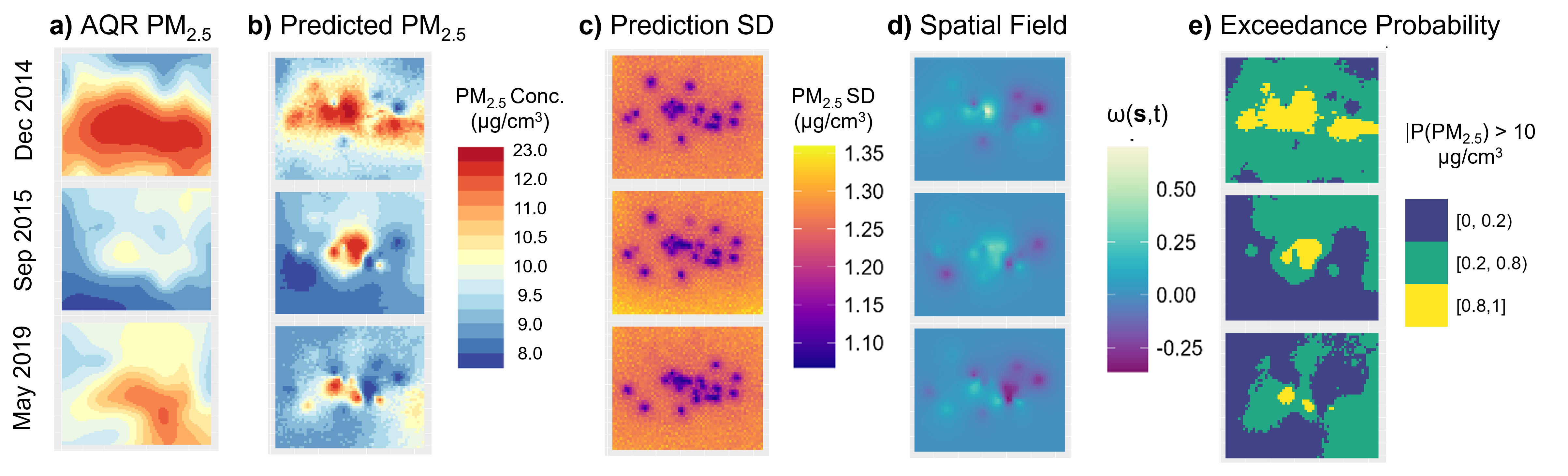}
    \caption{Plots of posterior modelling outputs for three example months: December 2014, September 2015 and May 2019. \textbf{a)} Modelled PM$_{2.5}$ concentration in $\mu g/m^3$ from the the UK AQR model. \textbf{b)} Mean posterior predicted PM$_{2.5}$ concentration in $\mu g/m^3$ from the best performing model. \textbf{c)} Standard deviation of the posterior predicted PM$_{2.5}$ concentration in $\mu g/m^3$. \textbf{d)} Mean posterior of the spatial random field for the spatiotemporal SPDE term. \textbf{e)} Exceedance probability: Probability of posterior predicted PM$_{2.5}$ concentration being greater than the 2021 WHO Air Quality Guideline interim target of 10 $\mu g/m^3$. \label{fig: Pred Plots}}
\end{figure}

Figure \ref{fig: Pred Plots} compares the Air Quality Reanalysis (AQR) model output (panel \textbf{a}) with our predicted PM$_{2.5}$ concentrations (posterior mean; panel \textbf{b}) for three randomly selected months, December 2014, September 2015 and May 2019. The predictions capture the main spatial patterns observed in the AQR data, including the elevated concentrations in central and eastern London.
Panel \textbf{c} shows the posterior standard deviation (SD), with higher uncertainty generally observed in peripheral areas, where monitoring coverage is sparser.
Panel \textbf{d} displays the posterior mean of the spatial random field from the SPDE model, representing structured residual spatial variation.

Finally, panel \textbf{e} presents the posterior probability that PM$_{2.5}$ concentrations exceed the WHO interim guideline of 10 $\mu$g/m$^3$. Exceedance probabilities are highest in central areas, particularly during the winter month (January).

\begin{figure}
    \centering
    \includegraphics[width=0.8\textwidth]{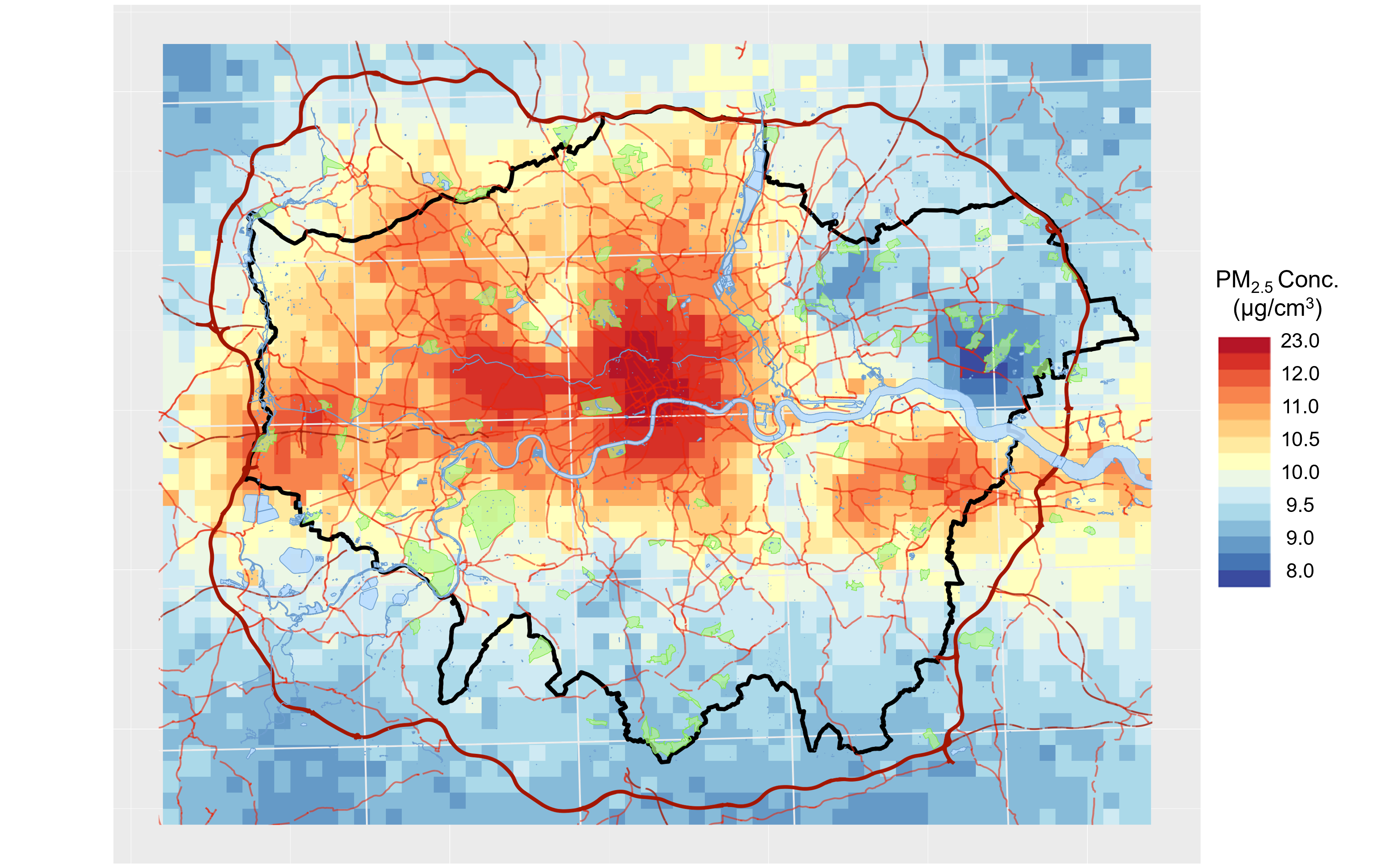}
    \caption{Mean posterior predicted PM2.5 concentration in $\mu$g/cm3 from the best-performing model for December 2014. Overlain by the map of the Greater London Authority Area from Figure 1 and with major roads and landmarks from OpenStreetMap. \label{Fig: Pred Map}}
\end{figure}

Table \ref{tab:Paras} presents the summary statistics of the marginal posterior distribution of the fixed and random model parameters and hyperparameters of the best , Model 4b: mean, standard deviation (SD), 2.5\%, 50\% and 97.5\% percentiles. 

The posterior mean of the model intercept, $\beta_0$, is 13.87 $\mu g/m^3$ ($\exp(2.63)$), and it represents the mean concentration of PM$_{2.5}$ across the study domain. As expected, the background indicator has a negative association with air particles and the 95\% CI does not include the null value of 0 (-0.263; 95\% CI(-0.319, -0.207)), suggesting that background monitoring sites and areas have lower pollution levels than traffic-labelled areas. Moreover, the PM$_{2.5}$ PCM model did not exhibit a clear directed effect in this model (-0.008; 95\% CI(-0.039, 0.023)); nonetheless, its inclusion does improve model performance. 

\begin{table}[p]
    \centering
    \small\addtolength{\tabcolsep}{5pt}

    \begin{tabular}{l l l l l l}
    \hline 
         & Mean & SD & 0.025 & Median & 0.975 \\ 
         \hline \hline
        $\beta_0 $ & 2.63 & 0.05 & 2.52 & 2.63 & 2.73 \\ 
        $\beta_{BG}$ & -0.27 & 0.03 & -0.33 & -0.27 & -0.21 \\ 
        $\beta_{PCM}$ & -0.008 & 0.02 & -0.04 & -0.008 & 0.02 \\ 

        \hline
        $\sigma_{\varepsilon}^2$ & 0.018 & 0.0008 & 0.017 & 0.018 & 0.020 \\ 
        $a$ & 0.95 & 0.01 & 0.92 & 0.95 & 0.97 \\ 
        $\sigma_{\omega}^2$ & 0.25 & 0.02 & 0.20 & 0.25 & 0.29 \\ 
        $\rho$ & 6.38 & 2.10 & 4.41 & 6.312 & 8.72 \\
        \hline
        $\sigma_{AQR}^2$ & 0.15 & 0.04 & 0.09 & 0.15 & 0.24 \\ 
        $\rho_{AQR}$ & 58.55 & 19.4 & 30.1 & 55.3 & 105.8 \\ 
        \hline
        $\sigma_{NDVI}^2$ & 0.63 & 0.18 & 0.37 & 0.60 & 1.06 \\
        $\phi_{NDVI}$ & 0.99 & 0.003 & 0.98 & 0.99 & 1.00 \\ 
    \hline \hline
    
    \end{tabular}
    \caption{Summary statistics of the posterior distribution of parameters and hyperparameters of the chosen model, \textbf{Model 4b}..\label{tab:Paras}}
\end{table}

The posterior estimates of these hyperparameters were similar in all competitive models, indicating good convergence of the modelling approach. The SPDE term has ranges such as: $a \in [0.94, 0.96]$ and $\sigma^2_{\omega} \in [0.23, 0.26]$. The SVC and TVC model terms also had some similar model estimates, depending on chosen covariates, e.g., for Models 4a - 4d: $\sigma^2_{\beta} \in [0.62,0.75]$ and $\phi_{\beta} \in [0.991, 0.999]$, but spatial range for the SVC varied ($\rho_{\beta} \in [51.2, 85.6]$). 

Based on the spatiotemporal SPDE specification in the final models (Equations 2-13), we estimated a spatial range of 5.8 km (95\% CI(4.3, 8.3)) and a high AR(1) coefficient ($a = 0.99$),  suggesting strong temporal autocorrelation. 

\begin{figure}
    \begin{subfigure}[b]{0.46\textwidth}
        \raggedleft
        \includegraphics[width=1\textwidth]{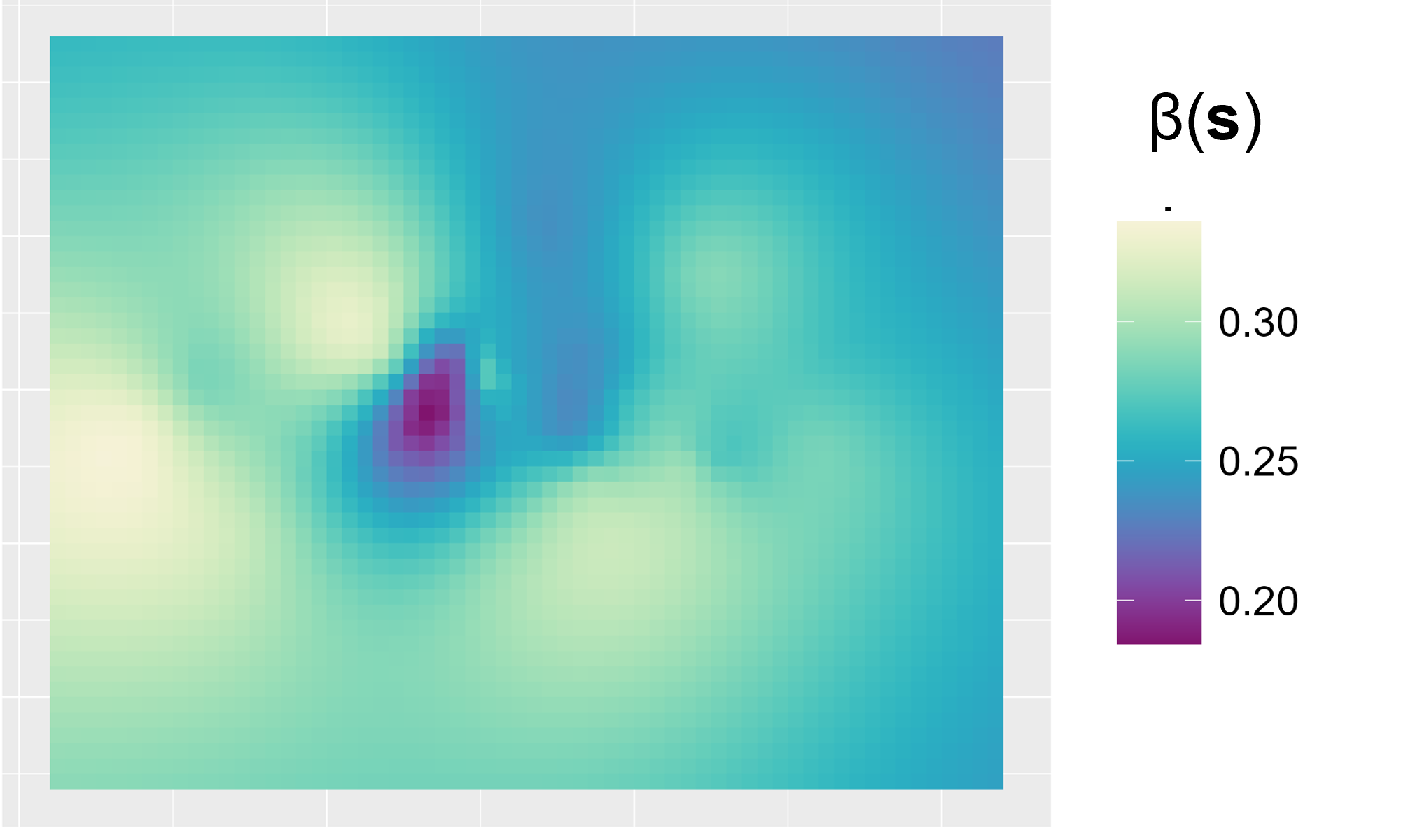}
         \caption{Mean posterior values of the spatial field for the SPDE SVC model on the AQR. \label{Fig: SVC}}
    \end{subfigure}
    \begin{subfigure}[b]{0.55\textwidth}
        \raggedleft
        \includegraphics[width=1\textwidth]{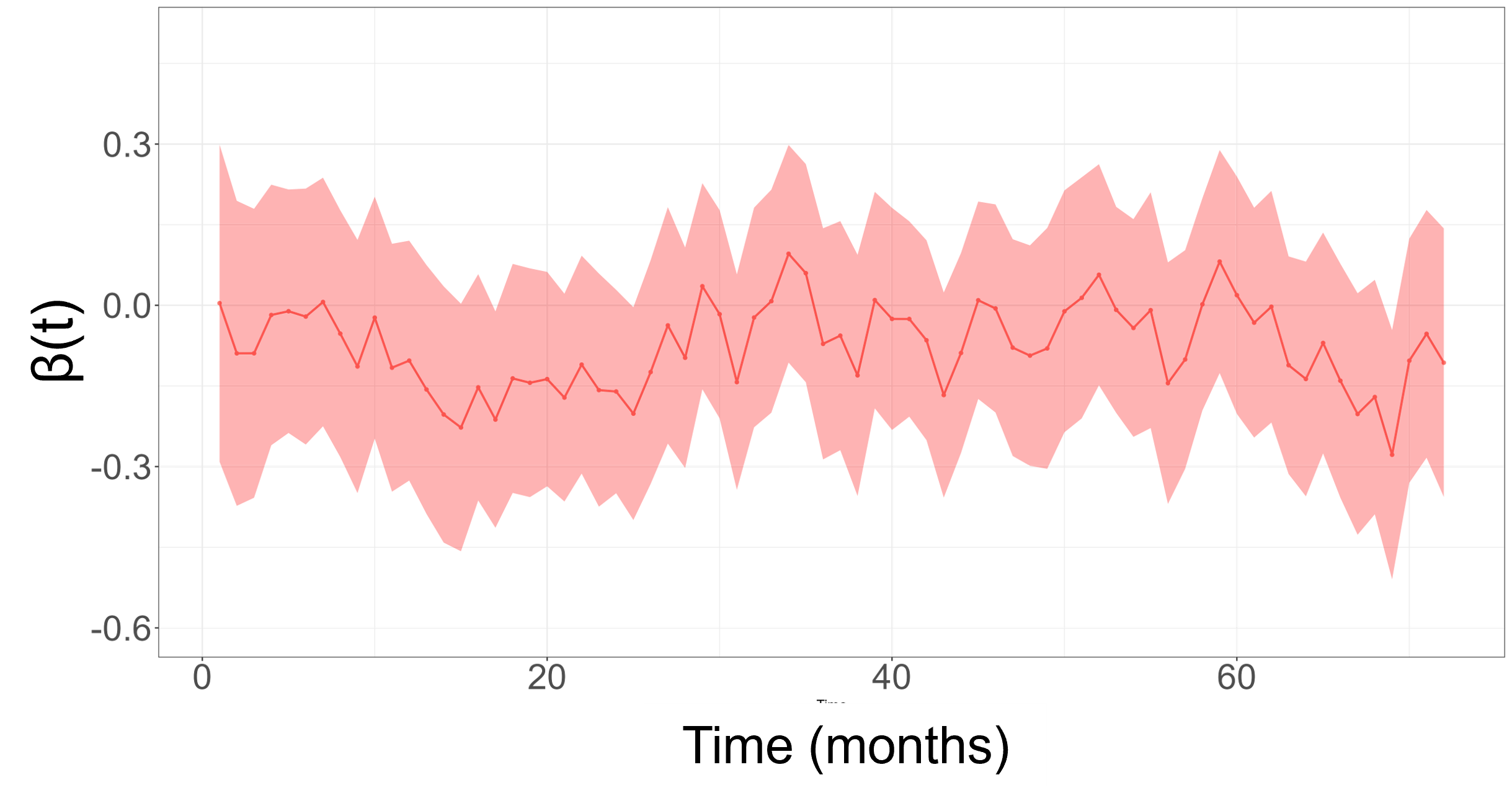}
        \caption{Mean posterior values of the temporally-varying coefficient on NDVI with 2.5\% and 97.5\% confidence intervals. \label{Fig: TVC}}
    \end{subfigure}
    \caption{Estimated spatially- and temporally-varying coefficient values\label{Fig: SVC and TVC}}
\end{figure}

Similarly, for the SVC model on the AQR variable, we used a spatial SPDE model (Equation \ref{Eq: SVC}). The spatial range was 67.2 km with a relatively large 95\% CI (35.6, 108.9). The mean estimated spatial field for this SVC SPDE model is plotted in Figure \ref{Fig: SVC}. For the TVC model on the NDVI variable, we specified an AR1 model. Hence, the posterior hyperparameter estimates for this model are $\sigma^2_\beta$ = 0.67, and $\phi_\beta$ = 0.99. The monthly posterior estimated values for $\beta_{NDVI}(t)$ are plotted in Figure \ref{Fig: TVC}, with 2.5\% and 97.5\% credible intervals.

\section{Discussion}\label{Section: Discussion}
This study developed and compared competitive spatiotemporal models for predicting urban air particle concentrations. By integrating multiple data sources, we demonstrated through robust cross-validation that our approach can effectively capture fine spatiotemporal patterns of air particles. We then produced accurate predicted concentrations throughout the study area. 

Our study integrated multiple data sources into a unified spatiotemporal model. As pointed out by \cite{Liverani2016ModellingData}, robust variable selection is difficult when working with spatial variables with high multicollinearity and spatial correlation. Our correlation analysis and covariate selection based on the model metrics and PMCC, as shown in the supplementary material, created a solid base for the development of our data fusion approach. 

One of the main strengths of our proposed method lies in its ability to reconstruct the entire latent field using a Bayesian framework. This approach enabled us to derive the marginal posterior distributions of all uncertainty parameters, thereby offering a rigorous quantification of the uncertainty in the particle concentration predictions, which is not achievable with non-model-based approaches.

We found clear seasonal patterns for PM$_{2.5}$, with higher levels in February-April and lower levels in the summer months, showing a slight overall negative relationship with the monthly mean temperature and a positive relationship for humidity and precipitation. These findings are in line with \cite{Liu2009EstimatingInformation}, which studied the geographical and physical processes behind spatiotemporal variability in PM$_{2.5}$ levels in the UK. However, as part of our correlation evaluations, we observed high correlations between many predictors, which informed our variable selection. Additionally, calculating monthly correlation between PM$_{2.5}$ and key variables showed strong seasonality, suggesting more complex non-linear processes \cite{Liu2009EstimatingInformation}.

Our final model accommodates non-linearity and non-stationarity in the underlying processes, consistent with the hypotheses proposed by \cite{Mukhopadhyay2018AWales}, through using varying-coefficient models.  

Despite the physical plausibility and promising results from previous studies, e.g., \cite{deHoogh2018ModellingSwitzerland}, we did not see a strong correlation between AOD and ground PM$_{2.5}$. This may be due to the latitude of Greater London, the high cloud cover, or the local climate; as discussed in \cite{Christopher2020GlobalRelationships}. Inclusion of just a linear effect term for AOD had little effect on model performance and the 95\% CI included the null value of 0. Moreover, applying spatially or temporally varying coefficients showed marginal improvements, which did not produce a strong competitive model. In contrast, NDVI has been shown previously to be an important environmental predictor of PM$_{2.5}$ \cite{Ai2023TheChina}. We saw evidence of this relationship between PM$_{2.5}$ and vegetation in our models, especially with an increase in performance when NDVI was included with a TVC term.

The PCM and AQR models were key drivers in all the competitive models. Integrating these data sources into the model brought together the high spatial-, low temporal-resolution of the PCM model and the low spatial-, high temporal-resolution of the AQR model. Each brought additional layers of information from emissions, meteorological data, and dispersion models. Moreover, other variables, such as temperature and wind, generally did not aid model performance and were not included in any final models. Due to the already incorporated local climate and emissions \cite{Savage2013AirEvaluation, Neal2014ApplicationForecast}, there may have been an issue of multicollinearity when additional variables are included.

Given the final model, the key interpretability comes from the hyperparameters in the INLA-SPDE construction and the coefficient varying models \cite{Bakka2018SpatialReview}. For the spatiotemporal SPDE, we estimated the spatial range ($\rho$ = 6.4), the standard deviation ($\sigma_{\omega}^2$ = 0.25) and the parameter of the AR1 model ($a$ = 0.95), see Table \ref{tab:Paras}. 

There is also an intrinsic link between the effective spatial range and the best practices for mesh construction. \cite{Belmont2022BuildingINLA} provides a comprehensive guide for 2d mesh construction based on the data. In our study, we implemented these suggestions and used the assessment tool laid out in \cite{Lindgren2018SpatiallyQuality} to select our mesh, weighing up model performance and complexity.

Figure \ref{Fig: TVC} shows that the TVC values for NDVI vary between -0.28 and 0.10, similar to the NDVI coefficient values in the covariate-only model, Model 1b. This non-linear and primarily negative relationship between vegetation indices and PM$_{2.5}$ is well replicated in literature \cite{Ai2023TheChina, Sheng2023SpatiotemporalStudy}. Also, there are physical processes that explain how urban vegetation reduces airborne particulate matter, through interception and absorption  \cite{Zhai2022StudyParticles, Nowak2006AirStates}.

Observing the predicted maps in Figure \ref{Fig: Pred Map}, we can visually identify regions of interest and specific times. 

Some prominent emission sources of PM$_{2.5}$ can be identified, especially dense road areas and the industrial areas along the East of the River Thames, around London City airport. Alternatively, there are identifiable air pollution sinks or areas of low emissions, including large parks and residential areas, and more rural areas around London.

In the time series plot (Figure \ref{Fig: Obvs vs Fit}), we can observe good adherence of the model to the observed site data at example sites, capturing the evolution over time and the seasonality. Another outstanding feature is the peak of predicted PM$_{2.5}$ in September 2014 (see in Figure \ref{Fig: Obvs vs Fit}; month 9). This is identified as an outlying month for climate, recorded as the driest September in UK data at the time \cite{Kendon2014State2014} and the warmest September globally since 1880 \cite{NCEI2014Monthly2014}. This likely drove the increase in PM$_{2.5}$ for this month, which matches the observed site data seen in Figure \ref{fig: Monitoring Sites} and the AQR model data.
 
\cite{Shen2024Monthly2019} presented a recent LUR approach to modelling monthly air pollution for Europe 2000 to 2019. Comparatively, our work is in line with the model performance of this study, i.e., for temporal CV on the $\mu g / m^3$: our R$^2$ = 0.67 is within their range  R$^2$ = 0.23 – 0.85 and, on the same scale, our RMSE = 2.8 within their range RMSE = 2.0 - 4.6. However, the Bayesian framework used in our study produces additional benefits. Through the Bayesian lens, we obtained the posterior predictive distributions for our model predictions, as well as measures of uncertainty in the model parameters and other outputs. This is key to interpreting the confidence in the model results. 

The Air Quality Standards Regulations 2010 \cite{UKGovernment2010The2010} set the Air Quality Guidelines (AQGs) for various pollutants and for PM$_{2.5}$ the annual mean concentration limit was set to be $20 \mu g /m ^3$. Most of the monitoring sites included in our study and over 99\% of all grid cells do achieve this goal annually, 2014 - 2019; however, London is falling short on the WHO Air Quality Guidelines 2021, set at just $5 \mu g /m^3$ for the annual mean target \cite{WorldHealthOrganization2021WHOGuidelines}. No monitoring sites nor grid cells achieved this goal, but around 42\% of grid areas in 2019 achieved the interim target of $10 \mu g /m ^3$.

The availability of the full posterior predictive distribution has also enabled us to produce exceedance probability maps, identifying areas with a high likelihood of exposure to adverse levels of air pollution. We adopted the 10 $\mu g / m^3$ threshold to evaluate the predicted monthly PM$_{2.5}$. This value reflects the interim WHO's air quality recommendations for annual PM$_{2.5}$, on the way towards the 5 $\mu g / m^3$, which is grounded in epidemiological evidence showing health risks even at low exposure levels. While not a regulatory limit on a monthly basis, applying this threshold provides a conservative and health-relevant benchmark to identify periods of potentially elevated exposure \cite{Papadogeorgou2019LowDirections}. Months exceeding 10 $\mu g / m^3$ can indeed indicate challenges in achieving compliance with the targets and highlight periods that may require targeted intervention.

Furthermore, the availability of the entire predictive distribution of PM$_{2.5}$ concentrations may inform any subsequent epidemiological studies evaluating the health effects of air pollution, allowing methodological robustness and accurate inference through propagation of the uncertainty from the exposure model to the health model, such as in \citet{Comess2022AStillbirth} and \citet{Huang2018MultivariateUncertainty}.

A similar Bayesian hierarchical modelling approach has been proposed by \cite{Fioravanti2021Spatio-temporalApproach}, which modelled daily PM$_{10}$ data across Italy. Our model shows comparative performance, despite some differences in the cross-validation methodology, and we show an improvement in the RMSE and bias, although a lower score for predictive correlation and coverage. The difference with our cross-validation methodology is due to the fact that the Authors broke up their spatiotemporal model into separate monthly models, therefore sacrificing an interpretable overall trend. This work is also related to previous modelling of NO$_2$ for London, seen in \cite{Forlani2020AR-INLA}, which also explores the use of the PCM and AQR. Methodologically, our advancement comes from using a newer reanalysis of the AQR model (\cite{Neal2014ApplicationForecast, MetOfficeUKReanalysis}) and the further SVC and TVC terms.

Addressing the change-of-support issues, we first carefully selected data sources and versions to match our intended model output, for easy integration. Data products were selected because they matched or complemented the 1km x 1km BNG grid and monthly resolution. Otherwise, the data was transformed, aggregated or interpolated, as seen in the data preparation code on GitHub. This spatial misalignment consideration is most influential for the AQR, which was only available at the larger 0.1 \textdegree resolution, which we successfully addressed by using an upscaling model framework with additional spatially-varying coefficient terms.

A limitation of this study is the difficulty with scalability in many ways. Firstly, London is a relatively small area and has a higher concentration of monitoring sites than other cities and rural areas. An extension of this model to all UK would mean a change in monitoring site density from roughly 40 km$^2$/site to >1,000 km$^2$/site with fewer locally managed networks, like London Air, too. Many studies explore the statistical challenges of having too few monitoring sites for accurate predictions (e.g., \cite{Fattoruso2020SiteMonitoring}) and others discuss solutions to this, like using low-cost monitoring \cite{Morawska2018ApplicationsGone}. 

This links to our proposal to use large, open-source and sometimes global datasets as predictive variables such as the AQR and PCM models, AOD, NDVI, and climate data. Here we have shown some predictive capabilities of these variables, especially through non-linear and more complex model terms. Hence, as these datasets are readily available across the UK and further afield, there is potential for similar modelling approaches to be applied in other urban contexts \cite{Christopher2020GlobalRelationships}.

Towards the SDGs on environment and health, global air pollution monitoring would be beneficial to identify priority locations and follow progress \cite{Moyer2019AreGoals, Ballari2023SatelliteApplications}. We can further use the results of this study to track the progress towards the more strict air quality recommendations by WHO \cite{WorldHealthOrganization2021WHOGuidelines} and to assess local air pollution policies, like the London Ultra Low Emissions Zone (ULEZ) \cite{GreaterLondonAuthority2024London-wideReport}.

Another issue with model scalability is resources and time. The final model requires high-performance computing resources and takes 5 hours to run. Any further model complexities, finer spatial resolution/mesh, or additional data would increase these demands. This can be the case with large INLA models, and there are various options, such as further parallelisation, splitting into the fitting and predicting of the model, or other programs to speed up algebraic computation \cite{AbdulFattah2025INLA+:HPC}.

The resulting air pollution predictions from this work is available on the GitHub platform (\url{github.com/abiril/AirPollutionModel}). Our estimates, with quantified uncertainty, can be used within health studies to estimate individual participants' or a population's exposure to fine particulate matter \cite{Han2017HumanPollution}. The fine spatial resolution and monthly estimates can provide highly specific exposure estimates for both mid-term to long-term exposures with any monthly lag \cite{Kumar2016ThePM10}.

\section{Conclusions}\label{Section: Conclusions}
In this paper, we have developed and compared Bayesian spatiotemporal hierarchical models for predicting monthly mean PM$_{2.5}$ concentrations in the highly urbanised setting of Greater London through a data fusion approach. 

The result is a full 1km x 1km grid of fine air particle predictions across Greater London for all months from 2014 to 2019, tested against different model formulations and other studies. The model demonstrates strong performance, achieving a good balance between complexity, computation speed, and predictive accuracy.

Using the INLA-SPDE framework, this work has tested and used a range of modern modelling methods, carefully considering the SPDE approach and additional model terms. We used a range of both local and global geospatial datasets, considering the predictive capabilities and inclusion through non-linear and non-stationary model components. 

Furthermore, the results from this work can be used in a wide range of further studies and applications. Directly, this work can be used for air quality surveillance for Greater London, as reported statistics, step-changes around new policies, and compliance with air pollution targets \cite{AirQualityExpertGroup2012FineKingdom}. Additionally, the model results can be used to quantify human exposure to outdoor air pollution in London, to be used in health studies \cite{Manisalidis2020EnvironmentalReview} and work towards the UN SDGs for 'promoting clean air and health' \cite{WorldHealthOrganization2021WHOMonoxide, WorldHealthOrganization2022AmbientPollution}.

Finally, we have contributed to both statistical methodology and environmental monitoring. We have brought together modern modelling ideas and datasets, while also balancing model complexity with predictive capability.  A key feature of this work is the transparency and reproducibility of the novel methods and code, as well as making the new estimates freely available, including the associated model uncertainty, giving the potential for new and more accurate health applications.

\section{Acknowledgements}

This study is a match-funded studentship by the School of Public Health, Imperial College, through the Medical Research Council (MRC) Centre in Environment and Health from Imperial College London, with Grant Reference number MR/T502613/1. All authors acknowledge Infrastructure support for the Department of Epidemiology and Biostatistics provided by the NIHR Imperial Biomedical Research Centre (BRC). 

\section{Supplementary Material}\label{Section: Sup Mat}
Additional information and supporting material for this article is available online at the journal's website and the GitHub repository: \url{github.com/abiril/AirPollutionModel}.

\bibliography{bibliography}

\end{document}